\documentclass[%
reprint,
superscriptaddress,
showpacs,
amsmath,amssymb,amsfonts,
aps,
prl,
floatfix,notitlepage,latexsym
]{revtex4-1}

\usepackage[dvipdfmx]{graphicx}
\usepackage{dcolumn}
\usepackage{bm}
\usepackage{braket}
\usepackage{mediabb} 
\usepackage[dvipdfmx]{attachfile2}

\begin{abstract}
Breaking the diffraction limit and focusing laser beams to subwavelength scale are becoming possible with the help of recent developments in plasmonics. Such subwavelength focusing bridges different length scales of laser beams and matter. Here we consider optical vortex, or laser beam carrying orbital angular momentum (OAM) and discuss potential subwavelength magnetic phenomena induced by such laser. On the basis of numerical calculations using Landau-Lifshitz-Gilbert equation, we propose two OAM-dependent phenomena induced by optical vortices, generation of radially anisotropic spin waves and generation of topological defects in chiral magnets. The former could lead to the transient topological Hall effect through the laser-induced scalar spin chirality, and the latter reduces the timescale of generating skyrmionic defects by several orders compared to other known means.

\vspace{3mm}
\noindent PhySH: Angular momentum of light, Ultrafast phenomena, Spin waves, Skyrmions, LLG

\end{abstract}

\begin{document}


\title{Encoding orbital angular momentum of light in magnets}
\author{Hiroyuki Fujita}
\thanks{Corresponding author}
\affiliation{Institute for Solid State Physics, University of Tokyo, Kashiwa 277-8581, Japan}
\affiliation{Kavli Institute for Theoretical Physics, University of California, Santa Barbara, California 93106, USA}
\email{h-fujita@issp.u-tokyo.ac.jp}
\author{Masahiro Sato}
\affiliation{Department of Physics, Ibaraki University, 
Mito, Ibaraki 310-8512, Japan}
\affiliation{Spin Quantum Rectification Project, ERATO, Japan Science and Technology Agency, Sendai
980-8577, Japan}
\email{masahiro.sato.phys@vc.ibaraki.ac.jp}
\date{\today}

\maketitle

{\it Introduction---}
Realization of ultrashort laser pulses with femto- to  pico- second width now offers a powerful tool for the study of non-equilibrium, ultrafast phenomena in solids. Since the pioneering observation of ultrafast demagnetization in Nickel by Beaurepaire et al in 1996~\cite{PhysRevLett.76.4250}, such highly non-equilibrium, laser-induced physics  is one of the most important research subjects in the field of magneto-optics~\cite{RevModPhys.82.2731,PhysRevLett.99.047601,Satoh:2012aa,PhysRevLett.110.177205,PhysRevB.88.220401,Mikhaylovskiy:2015aa,R.:2016aa,arXiv:1404.2010,PhysRevB.90.085150,PhysRevB.90.214413,Mentink:2015aa,Pimenov:2006aa,PhysRevLett.105.147202,Takahashi:2012aa,arXiv:1602.03702}. 

So far, most of ultrafast magneto-optical phenomena in solids are explored using laser beams with Gaussian spatial profile. In 1992, however, Allen {\it et al.} proposed~\cite{PhysRevA.45.8185} a new type of laser beam, now called optical vortex~\cite{9780511795213}. Optical vortex is a beam carrying orbital angular momentum (OAM), which can be transferred to physical systems as, for example, mechanical rotation of (semi-) classical particles~\cite{Friese:1998aa,PhysRevLett.75.826,Paterson:2001aa, PhysRevB.93.045205}. In the past decades, many applications such as super-resolution microscope~\cite{PhysRevLett.98.218103} and optical ablation~\cite{Toyoda:2012aa,Hamazaki:10} are developed. However, the use of optical vortex for controlling microscopic magnetic degrees of freedom of solids is almost unexplored. 

The main difficulty in is in the mismatch between the timescales. Typically, the timescale of spin dynamics is in the Tera Hz (THz) or Giga Hz region, but the wavelength of optical vortex of such frequencies is too large for individual spins  to ``feel" its characteristic spatial profile. Using heating effect, as discussed in Ref.~\onlinecite{Fujita2016} is one option to resolve this issue, but here we take another approach: breaking the diffraction limit~\cite{:/content/aip/journal/jap/98/1/10.1063/1.1951057,PhysRevLett.92.143904,Huang:2007aa,Gramotnev:2010aa} of THz beams to realize the subwavelencth optical vortices~\cite{Heeres:2014aa}. Stimulated by the enormous successes for the microwave and visible light, breaking the diffraction limit of THz beams is actively explored and the technologies are rapidly developing recently with the help of plasmonics. We can now design the spatial profile of THz beams at the subwavelength scale~\cite{Tanaka_2016} and actually focus the THz optical vortices~\cite{Arikawa:17} using designed metallic structures such as an array of antennas. Although the technology is still primitive, the prospect for the deep subwavelength focusing of THz beams stimulates studies of ultrafast magnetic phenomena induced by such beams. 

In this paper, with numerical calculations based on the Landau-Lifshitz-Gilbert (LLG) equation, we seek the ultrafast manipulation of magnets through Zeeman and magneto-electric (ME) coupling between spins and the subwavelength optical vortices. We find that optical vortex can be used to excite multipolar and spiral spin waves with OAM dependent wavefronts, which enable us to dynamically generate inhomogeneous spin texture and would induce the transient topological Hall effect~\cite{PhysRevLett.83.3737,PhysRevB.62.R6065,Taguchi:2001aa,PhysRevLett.87.116801,:2002aa,PhysRevLett.102.186602}.
 
Moreover, for chiral ferromagnets with Dzyaloshinskii-Moriya (DM) interaction~\cite{DZYALOSHINSKY1958241,PhysRev.120.91} we observe OAM-dependent generation of topological magnetic defects such as skyrmioniums~\cite{Hubert1994, PhysRevLett.110.177205} and skyrmions~\cite{Bogdanov2006,Muhlbauer:2009aa,Fert:2013aa,Seki_BOOK}, both of which are prospected as ingredients of future magnetic memory devices. 
We find that OAM of lasers can be transferred as the topological number, i.e. the number of skyrmions generated by the beam. We show that the spatial profile of the subwavelength optical vortex offers an ideal tool for creating skyrmionic defects and the timescale of their generation can be orders of magnitude shorter than other known schemes like heating~\cite{Koshibae:2014aa,Fujita2016} and electric current pulses~\cite{Yuan:2016aa,PhysRevB.94.094420}.

{\it Optical vortex---}
Optical vortex, or Laguerre-Gaussian (LG) mode is a class of solutions of Maxwell's equations in a vacuum under the paraxial approximation. The derivation of LG modes can be found in literatures~\cite{PhysRevA.45.8185,9780511795213}. In the cylindrical coordinate $(\rho, \phi, z)$, where $\rho$ is the radial coordinate and $\phi$ the azimuthal angle, the field configuration of LG modes propagating in the $z$ direction is given as $\vec{B}(\rho, \phi, z = 0) \propto \vec{e}_p B_{m,p}(\rho, \phi, z =0)$ at the focal plane ($z=0)$. Here $\vec{e}_p$ is the polarization vector, $\vec{e}_p = \hat{x}, \hat{y}$ for linearly polarized light and $\vec{e}_p = \hat{x} \pm i\hat{y}$ for circularly polarized light. The spatial profile $B_{p,m}$ is characterized by two integers, radial index $p$ and OAM $m$:
\begin{align}
B_{m,p}(\rho, \phi,0) = \left(\frac{\rho }{W}\right)^{\left| m\right| } e^{-\frac{\rho ^2}{W^2}+i m \phi } L_p^{\left| m\right| }\left(\frac{2 \rho ^2}{W^2}\right),
\label{vortexbeam}
\end{align}
where $L_p^{|m|}(\cdot)$ is the generalized Laguerre function.

\begin{figure}[htbp]
\centering
\includegraphics[width = 85mm]{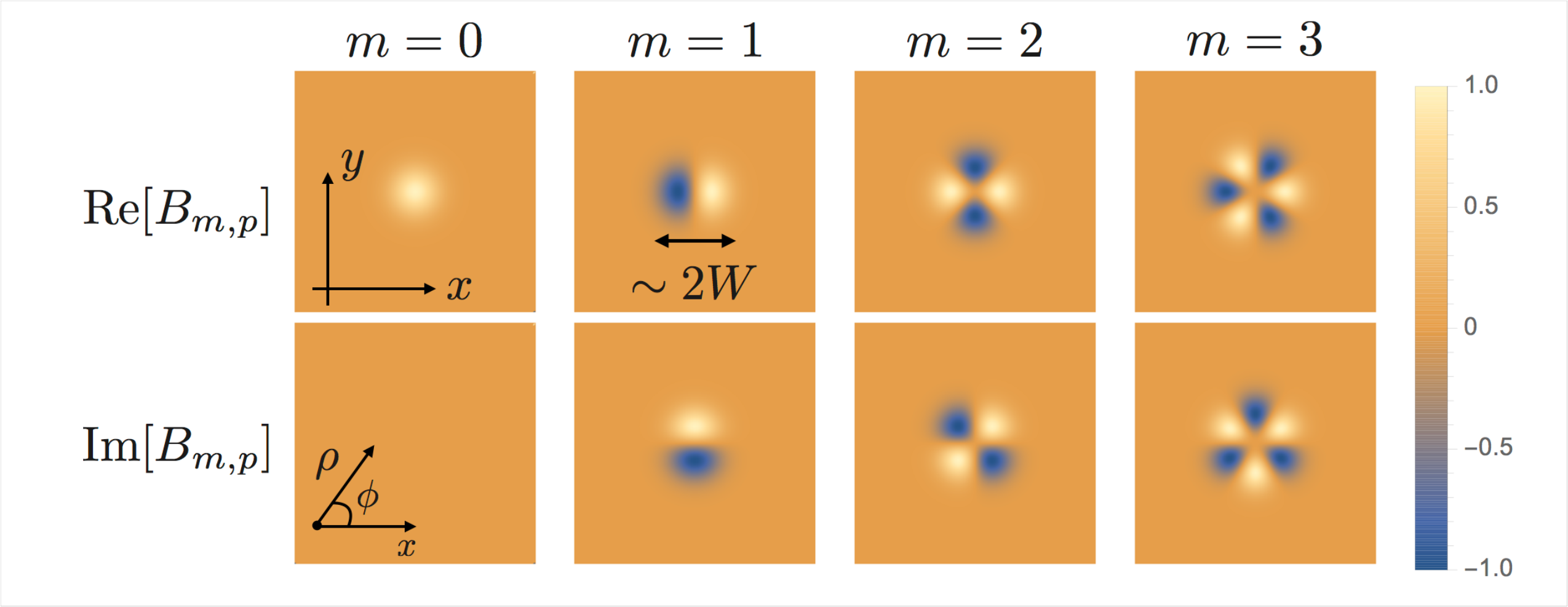}
       \caption{Snapshots of the spatial profile of magnetic fields of the optical vortex Eq.~\eqref{vortexbeam} (for $p = 0$) at the focal plane ($z = 0$) for several values of OAM. For a beam with OAM $m$, if we go around the topological singularity $\rho = 0$, the magnetic field changes its sign $2m$ times. The peak values of the fields are normalized to unity.}
          \label{distribution} 
  \end{figure}

 The non-vanishing phase twist for $m\neq 0$ requires the field to vanish at the topological singularity $\rho = 0$. The beam waist $W$ represents the size of optical vortex (see Fig.~\ref{distribution}), and usually cannot be smaller than a half the wavelength because of the diffraction limit. However, as we mentioned, by using plasmonics techniques~\cite{:/content/aip/journal/jap/98/1/10.1063/1.1951057,PhysRevLett.92.143904,Huang:2007aa,Gramotnev:2010aa,Heeres:2014aa,Tanaka_2016}, in principle we can take $W$ to be much smaller than the wavelength. In the rest of this paper, we consider the simplest optical vortex with the radial index $p=0$. We note that LG modes of electric fields, which we consider in the supplementary material, have the same form as magnetic fields presented above, as far as we are discussing the beams propagating in a vacuum.
  
{\it Model and numerical method---}
In this paper, we numerically investigate the laser-driven dynamics of (chiral) ferromagnets. Particularly, in the following, we focus on Zeeman coupling between spins and magnetic fields of optical vortices and study the spin dynamics in the framework of LLG equation. The effect of ME coupling, which can be important and useful in multiferroic materials~\cite{Cheong:2007aa,0034-4885-77-7-076501}, is discussed in the supplementary material, where we observe qualitatively the same results obtained for Zeeman coupling presented below. Depending on the sign of $m$, these magnetic field distributions rotate in either clockwise (CW) or counter-clockwise (CCW) way. As we discuss below, the spatially-inhomogeneous in-plane structure of optical vortices shown in Fig.~\ref{distribution} and its time-dependent rotation induce various characteristic phenomena. 
  
Let us consider the situation where a square lattice classical (chiral) ferromagnet is placed at the focal plane of the optical vortex. The Hamiltonian we consider is
\begin{align}
H &= - J \sum_{\vec{r}} \vec{m}_{\vec{r}}\cdot\left(\vec{m}_{\vec{r} + a \vec{e}_{x}}+ \vec{m}_{\vec{r} + a \vec{e}_{y}} \right)-H_z\sum_{\vec{r}}m^{z}_{\vec{r}} \nonumber \\
&+\sum_{\vec{r},i}\vec{D}_{i}\cdot\left(	\vec{m}_{\vec{r}}\times \vec{m}_{\vec{r}+a\vec{e}_{i}}		 \right) -\sum_{i} \vec{B}(\vec{r},t) \cdot\vec{m}_{\vec{r}},
\label{model}
\end{align}
where $a$ is the lattice constant and $\vec{e}_{i}$ is the unit vector along the $i$-axis ($i = x, y$). The vector $\vec{m}_{\vec{r}}$ represents the spin at the site $\vec{r}$, with its norm normalized to unity. We have the ferromagnetic Heisenberg interaction $J > 0$, DM interaction with DM vector $\vec{D}_{i} = D \vec{e}_{i}$ on the bond $(\vec{r}, \vec{r}+a\vec{e}_{i})$~\footnote{With this choice of DM vectors, we obtain so-called Bloch-type skyrmions.}, and the static external magnetic field applied in the $z$-direction $H_{z}$ aside from the optical vortex. The last term describes Zeeman coupling between spins and the optical vortex. 

The Hamiltonian Eq.~\eqref{model} is a canonical model of (chiral) ferromagnets. Despite its simplicity, this model well describes some experimental results of the actual three dimensional (thin film) materials and is widely used for the study of their topological defects, i.e. skyrmions~\cite{Bogdanov2006,Muhlbauer:2009aa,Fert:2013aa,Seki_BOOK} (see Ref.~\onlinecite{Seki_BOOK} for the review). With increasing the external field $H_z$, the model shows the helical ordered phase, skyrmion lattice phase, and the ferromagnetic phase as observed in thin film materials~\cite{Yu:2010aa}. In the supplementary material, we give a brief review of this model, the phase diagram and topological defects therein. Depending on the materials, the size of skyrmions can vary from nm to $\mu$m~\cite{Fert:2013aa,Seki_BOOK,ShibataK.:2013aa}.  

The dynamics of spins under the applied optical vortex is determined by the following LLG equation for the model Eq.~\eqref{model}~\cite{Seki_BOOK}:
\begin{align}
\frac{d\vec{m}_{\vec{r}}}{dt} &= - \vec{m}_{\vec{r}} \times \vec{H}_{\mathrm{eff}}+ \alpha\vec{m}_{\vec{r}}\times \frac{d\vec{m}_{\vec{r}}}{dt}.
\label{LLGS}
\end{align}
The time coordinate $t$ is normalized by $\hbar/J$, which corresponds to $0.13$ ps for $J = 5$ meV. The second term in the right hand side of Eq.~\eqref{LLGS} is the Gilbert damping term describing the dissipation with strength determined by the dimensionless parameter $\alpha$. We see that each spin precesses around the normalized effective field $\vec{H}_{\mathrm{eff}} =-\vec{\nabla}_{\vec{m}_{\vec{r}}}(H/J)$, and damps towards that. Hereafter, we take $\hbar = J = 1$. Namely, in the following $H_z$, $\vec{B}$, and $\vec{D}$ are measured in the unit of $J$, and the time is in the unit of $\hbar/J$. 

In the THz region, heating caused by the laser absorption is small, so that we ignore the laser heating effect in the following. Actually, magnetic resonance experiments on magnets and multiferroic materials for THz light can be well explained by theory without taking the heating effect into account (for example, Refs~\onlinecite{1367-2630-18-1-013045,PhysRevLett.104.177206}). In Ref.~\onlinecite{1367-2630-18-1-013045}, for an antiferromagnetic dielectric HoFeO${}_3$ the temperature change caused by a short intense ($\sim$ 1 Tesla) THz magnetic field pulse at the magnetic resonance is estimated to be about 1mK ($< 10^{-4}$ meV). This is orders of magnitude smaller than the energy scale of the direct coupling between spins and light. 
\par
  
For the numerical calculations, we use the fourth-order Runge-Kutta method with numerical time step of the calculation $\Delta t = 0.2 \hbar/J$. We consider a system consisting of 150 times 150 sites with periodic boundary condition imposed in both $x$ and $y$ directions. In all the cases below, we assume a simple pulse of optical vortex: 
 \begin{align}
 B(\vec{r}, t) = \frac{|B_0|B_{m, p}(\vec{r})}{\max_{\vec{r}}|B_{m, p}(\vec{r})|}\exp \left[-\left(\frac{t - t_0}{\sigma}\right)^2 - i \omega t \right]  \label{magfield}
 \end{align}
 where $\omega$ is the frequency, $|B_0|$ determines the strength of the magnetic field, and $\sigma$ gives the beam duration. For $J \sim 5$ meV, $B_{0} = 0.1$ is about $9 $ Tesla, $\sigma = 20 \hbar/J$ about 3 ps, and $\omega = 1$ about 1.2 THz. In the following, we propose two ultrafast applications of optical
vortex, anisotropic spin wave excitations and generation of topological defects. In both cases, below we assume THz optical vortices with nanoscale beam waist $W\sim10a$, which is orders of magnitude smaller than the wavelength. As we show in the supplementary material and discuss lat
er, even with much larger beam waists, qualitatively the same results can be observed by using proper materials.

{\it Spin waves and magnetic resonance---}
First we apply optical vortex to ferromagnets with $D = 0$. In this case the sign of OAM is unimportant, but  the spatial profile of optical vortices still leads to spin wave excitations with characteristic spatial distribution depending on the value of OAM. Here we only focus on linearly polarized waves $\vec{e}_p = \hat{x}$ with finite OAM. In the supplementary material, we give a discussion for circularly polarized optical vortices and Gaussian beams without OAM.

\begin{figure}[htbp]
   \centering
\includegraphics[width = 88mm]{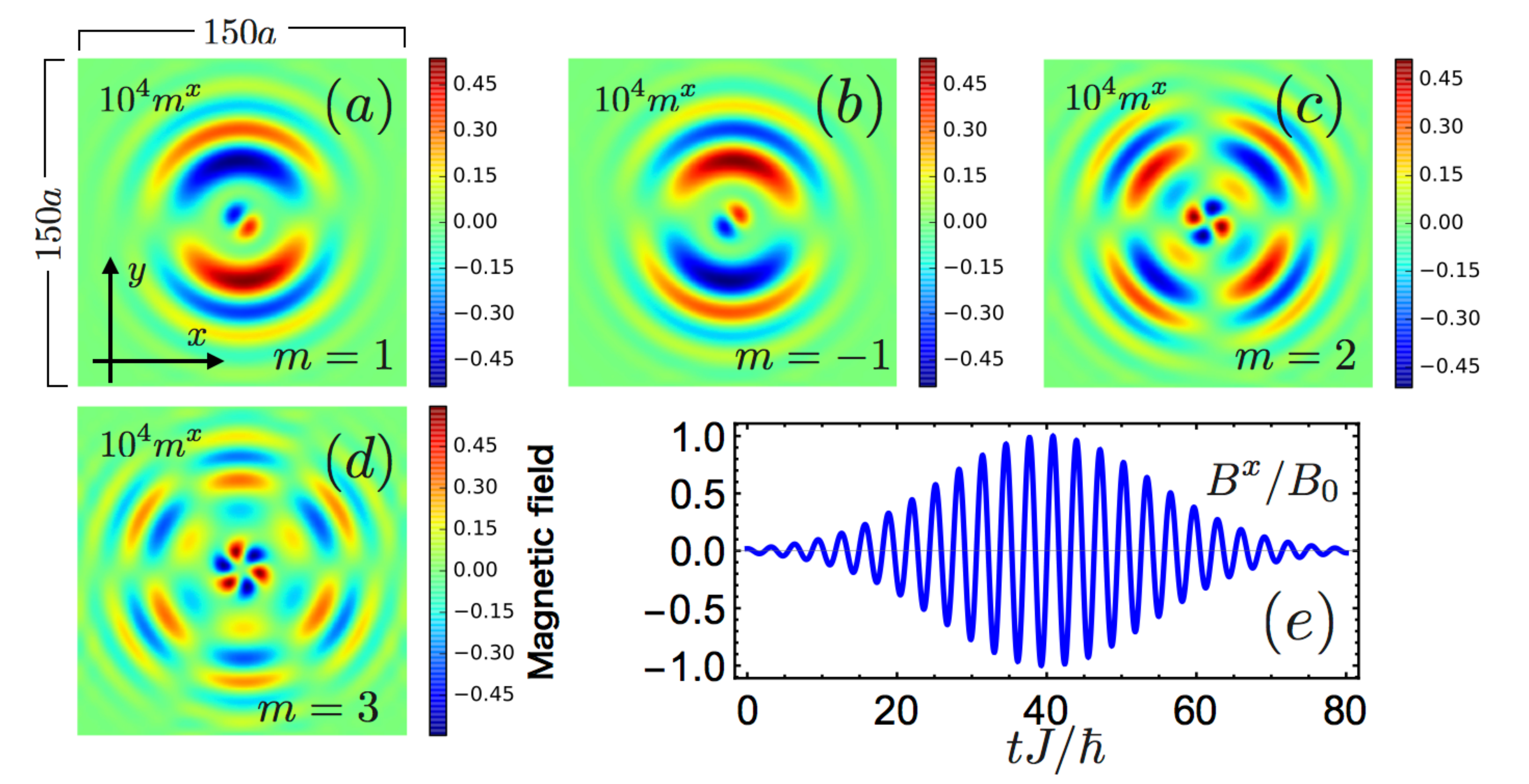}
       \caption{Multipolar spin wave radiation $(a$)-$(d)$ induced by linearly polarized optical vortices with $\vec{e}_{p} = \hat{x}$ for $D = 0$, $H_{z} = 0.015$, $W= 7.5 a$, $\omega = 2$, $\sigma = 20$, $t_0 = 40$, $B_0 = 0.05$, and $\alpha = 0.1$. We show the $x$-component of spins ($\times 10^4$) at $t = 80$ and the temporal profile of the magnetic field $(e)$ for $m = 1$ at $\rho = w/\sqrt{2}$ and $\phi = 0$. The initial state at $t = 0$ is the ferromagnetic state ($m^{z}_{\vec{r}} = 1$ for all sites $\vec{r}$). For $J = 5$ meV, $\omega = 2$ corresponds to $2.4$ THz and the beam amplitude $B_0 = 0.05$ does 4.3 Tesla.}
          \label{multipolar} 
  \end{figure}

\begin{figure}[htbp]
   \centering
\includegraphics[width = 90mm]{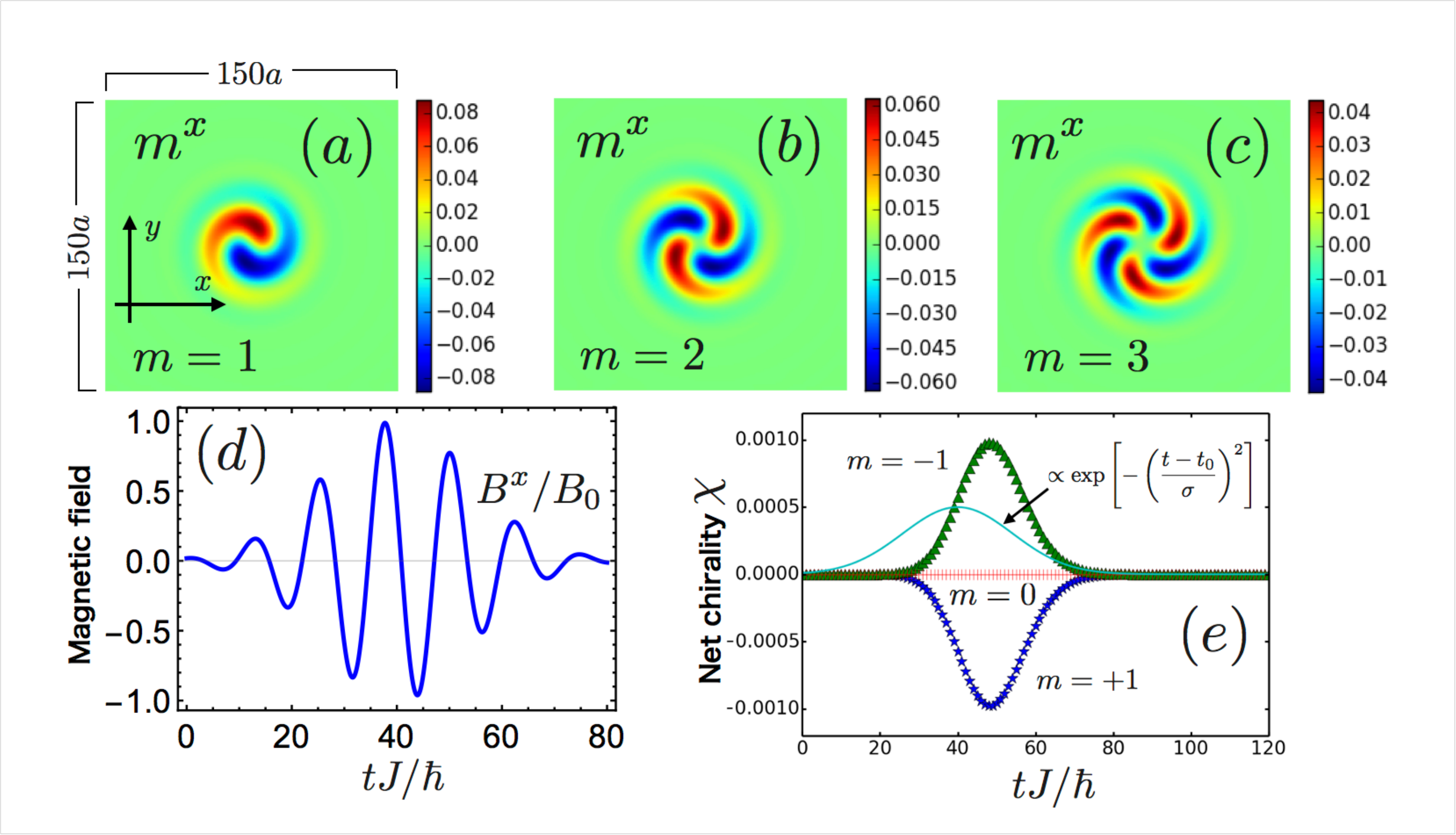}
       \caption{Spiral spin wave radiation induced by linearly polarized optical vortices $(a)$-$(c)$ at the magnetic resonance $H_{z} = \omega = 0.3$ (other parameters are the same as Fig.~\ref{multipolar}) and the dynamically induced net scalar spin chirality for $m = 0, \pm1$ $(e)$. Due to the anisotropic spin wave structure, we observe non-vanishing, OAM dependent net scalar spin chirality for $m\neq 0$. We also present the temporal profile of the field $(d)$ for $m = 1$ at $\rho = w/\sqrt{2}$ and $\phi = 0$. For $J = 5$ meV, $\omega = H_z = 0.3$ correspond to $0.4$ THz and $26$ Tesla.}
          \label{resonance} 
  \end{figure}

For high frequency beams satisfying $\omega \agt J, H_z$, as shown in Fig.~\ref{multipolar}, we have multipolar spin waves (dipolar, quadrupolar, octapolar) depending on OAM of those beams. On the other hand, at the magnetic resonance ($\omega \sim H_z$), the spin wave amplitudes become drastically larger, and the multipolar wavefronts connect with each other to be spiral-shaped as shown in Fig.~\ref{resonance}. 
In both cases, the spin structure is modulated from the collinear ferromagnetic state in an inhomogeneous way. Therefore, if (and only if) the laser beam carries OAM, we can dynamically induce the scalar spin chirality $\chi_{i,j,k} = \bm{S}_i\cdot(\bm{S}_j\times\bm{S}_k)$ ($j, k$ are neighboring sites of the site $i$) as shown in the right-bottom panel of Fig.~\ref{resonance}. The non-vanishing net chirality $\chi = \sum_i  \chi_{i,i+\hat{y},i+\hat{x}}+\chi_{i,i+\hat{x},i-\hat{y}}+\chi_{i,i-\hat{y},i-\hat{x}}+\chi_{i,i-\hat{x},i+\hat{y}}$ would lead to the topological Hall effect~\cite{PhysRevLett.83.3737,PhysRevB.62.R6065,Taguchi:2001aa,PhysRevLett.87.116801,:2002aa,PhysRevLett.102.186602} in itinerant magnets, but the quantitative analysis is beyond the scope of this paper and may be presented elsewhere.

{\it Generation of topological defects---}
As we see in Fig.~\ref{distribution}, the optical vortex with non-vanishing OAM has radially anisotropic field distribution. This induces either chiral or anti-chiral twist to the spin texture depending on the sign of the OAM. In chiral magnets with $D\neq0$, such twisted nature of the perturbation should compete with the intrinsic chirality of the magnets determined by their DM interaction.

\begin{figure}[htbp]
\centering

\includegraphics[width = 160mm]{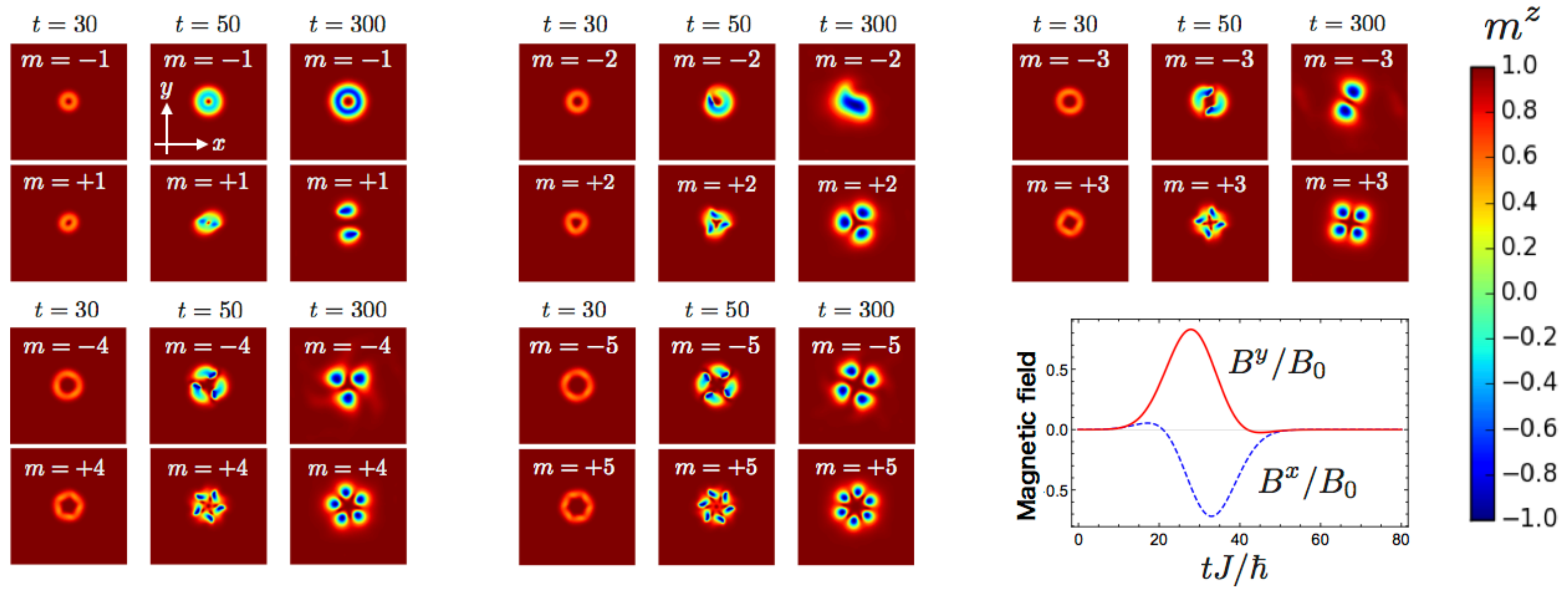}
       \caption{Orbital angular momentum dependent responses of a chiral ferromagnet against left-handed beams with $\vec{e}_{p} = \hat{x} + i \hat{y}$. The initial state at $t = 0$ is the ferromagnetic state where all spins point to the $+\hat{z}$ direction. For $D = 0.15$, $H_{z} = 0.015$, $W = 10 a$, $\omega = 0.075$, $\sigma = 10, t_0 = 30$, $B_0 = 0.15$, and $\alpha = 0.1$, we show time evolutions of spins for various orbital angular momentum $m $. We also show the temporal profile of the magnetic fields at $\rho = w/\sqrt{2}$ and $\phi = 0$ for $m=1$. The system size is 150$a$ times 150$a$ with periodic boundary. The ring-shaped object observed in the $m = -1$ case is a skyrmionium and point-like objects in other cases are skyrmions.}
          \label{chiral_dep} 
  \end{figure}

In the following calculations, we take left-handed optical vortices with polarization vector $\vec{e}_{p} = \hat{x} + i \hat{y}$. The polarization dependence and the case for Gaussian beams without OAM are discussed in the supplementary material. For left-handed beams, the twist induced to the spin texture by them is CW (CCW) for negative (positive) OAM. Since our model Eq.~\eqref{model} stabilizes skyrmions with CW spin twisting, for negative OAM, the chiralities of light and magnets are in a sense consistent. Here we take the following parameters: $W = 10 a$, $\omega = 0.075$, $\sigma = 10$, $t_0 = 30$, $B_0 = 0.15$, $D = 0.15$, $H_{z} = 0.015$, and $\alpha = 0.1$~\footnote{We confirm that distinct behaviors between positive and negative OAM is not an artifact of fine-tuning by performing calculations with slightly different beam profiles.}. With these beam parameters, the magnetic field Eq.~\eqref{magfield} becomes a half-cycle pulse both for $B^x$ and $B^y$ as shown in Fig.~\ref{chiral_dep}, and the beam waist is comparable with the size of skyrmions. For $J \sim 5$ meV, the frequency is about $0.1$ THz and the peak values of the fields are about $10 $ Tesla. We assume the initial state at $t=0$ to be the meta-stable perfect ferromagnetic state in the skyrmion crystal phase of the model Eq.~\eqref{model}. In this phase skyrmionic defects are once formed, stable against weak perturbations such as thermal fluctuations~\cite{Hubert1994} (experimentally skyrmions in some materials can survive even at the room temperature). Hence, even if we take the heating effect ignored in this paper into account, the following results will hold at least qualitatively.

In Fig.~\ref{chiral_dep}, we present the OAM dependence of the dynamics of the laser-irradiated chiral magnet. As we noted, in the present setup, $m < 0$ optical vortex twists the spin texture in a way consistent with the intrinsic chirality of the target. In combination with the topological singularity ($\rho=0$) where the field amplitude is zero, this results in the generation of skyrmioniums for $m = -1$. Due to the destructive effect of the frequent changes in the sign of magnetic fields around the topological singularity, for $m < -1$ we do not obtain skyrmionium. Nevertheless, the OAM dependence of the outcome is clear: the number of skyrmions after the irradiation is given by ${\rm sign}(m)(m+1)$. Therefore, we can encoded OAM of optical vortices into chiral magnets as their topological charge. Although the field strength assumed here is strong as THz beams, the field enhancement accompanied by the subwavelength focusing~\cite{Gramotnev:2010aa} would resolve this issue.

\onecolumngrid

\twocolumngrid

Here we comment on the timescale of the process we discussed. As for the creation of skyrmionium, the timescale of its creation with the present scheme (with $m = -1$ beam) turns out to be much shorter than other schemes, and is essentially unchanged  even the size of skyrmions is varied (see the supplementary material). The scheme using heating with vortex beams~\cite{Fujita2016} requires beams with period O$(100) \hbar/J$ long and that using spin-polarized current~\cite{PhysRevB.94.094420} needs a pulse with duration of O$(10^4) \hbar/J$~\footnote{They apply spin-polarized current of a few hundred pico seconds for the exchange coupling $15$ pJ$/$m, which would correspond to $J = 10 -100$ meV.}, while with our scheme its creation completes within O$(10) \hbar/J$ (see more details in the supplementary material). Behind the short timescale, there are two features of optical vortices: coherent coupling with spins and the $\phi$-dependent spatial profile, which generate the desired twisted spin texture directly in the THz timescale.

{\it Concluding remarks--}
In this paper, we proposed two ultrafast magnetic phenomena induced by subwavelength optical vortices. We found that OAM of optical vortex can be encoded in magnets in the form of anisotropic spin waves or topological defects. 

We show that there appears the non-vanishing net spin chirality due to the anisotropic spin waves, which would lead to the laser-induced topological Hall effect. With regard to topological defects in chiral magnets our findings offer a  scheme for the ultrafast generation of them. Unlike other known schemes, our method can generate multiple skyrmions at the same time in a controlled way. 

Finally, we comment on the experimental feasibility of the proposed phenomena. First, the excitation of the anisotropic spin waves in Figs.~2 and 3 are rather easy, since they are long-wavelength phenomena by nature and do not actually require the subwavelength focusing. On the other hand, the feasibility of the generation of topological defects in Fig.~\ref{chiral_dep} is more subtle. As we mentioned in the introductory part, the THz focusing is at the very early stage of its study, and at present the maximum focusing achieved experimentally is by a factor of three to four~\cite{Arikawa:17}, and it is indeed quite challenging to realize the ``nanometre" scale focusing we assumed in Fig.~\ref{chiral_dep}. However, as discussed in the supplementary material, we can verity that the proper beam waist for the proposed phenomena simply scales with the intrinsic length scale of the target materials, the size of skyrmions. Therefore, by using materials with large skyrmions discovered recently~\cite{ShibataK.:2013aa,Jiang283,Yu_2016,Woo:2016aa,Moreau-LuchaireC.:2016aa,Boulle:2016aa,Jiang:2016aa}, the requirement for the focusing factor can be drastically relaxed and we could realize the OAM encoding in Fig.~\ref{chiral_dep}. 

{\it Acknowledgement---}
We thank Koichiro Tanaka for useful comments. H. F. is supported by Advanced Leading Graduate Course for Photon Science (ALPS) of Japan Society for the Promotion of Science (JSPS) and JSPS KAKENHI Grant-in-Aid for JSPS Fellows Grant No.~JP16J04752. M. S. was supported by Grant-in-Aid for Scientific Research on Innovative Area, ”Nano Spin Conversion Science” (Grant No.17H05174), and JSPS KAKENHI (Grant No. JP17K05513 and No. JP15H02117). 
 This research was supported in part by the National Science Foundation under Grant No. NSF PHY-1125915.

\bibliography{EM}

\newpage

\onecolumngrid
\section{Supplementally Material for \\``Encoding orbital angular momentum of light in magnets"}

As in the main text, in this supplementary material we set $\hbar = J = 1$, namely measure energy in the unit of $J$ and time in $\hbar/J$. For all the numerical calculations based on the Landau-Lifshiz-Gilbert (LLG) equation below, we use the forth-order Runge-Kutta method with timestep $\Delta t = 0.2$ in spatially periodic systems. When $ J = 5$ meV, $\Delta t$ corresponds to 26 fs. The Gilbert damping constant is $\alpha = 0.1$, same as in the main text.
\setcounter{equation}{0}
\setcounter{section}{0}
\setcounter{figure}{0}
\setcounter{table}{0}
\setcounter{page}{1}
\makeatletter
\renewcommand{\theequation}{S\arabic{equation}}
\renewcommand{\thefigure}{S\arabic{figure}}
\renewcommand{\bibnumfmt}[1]{[S#1]}
\renewcommand{\citenumfont}[1]{S#1}
\section{Two dimensional chiral magnets and skyrmions} \label{Chiral}
\begin{figure}[htbp]
   \centering
\includegraphics[width = 140mm]{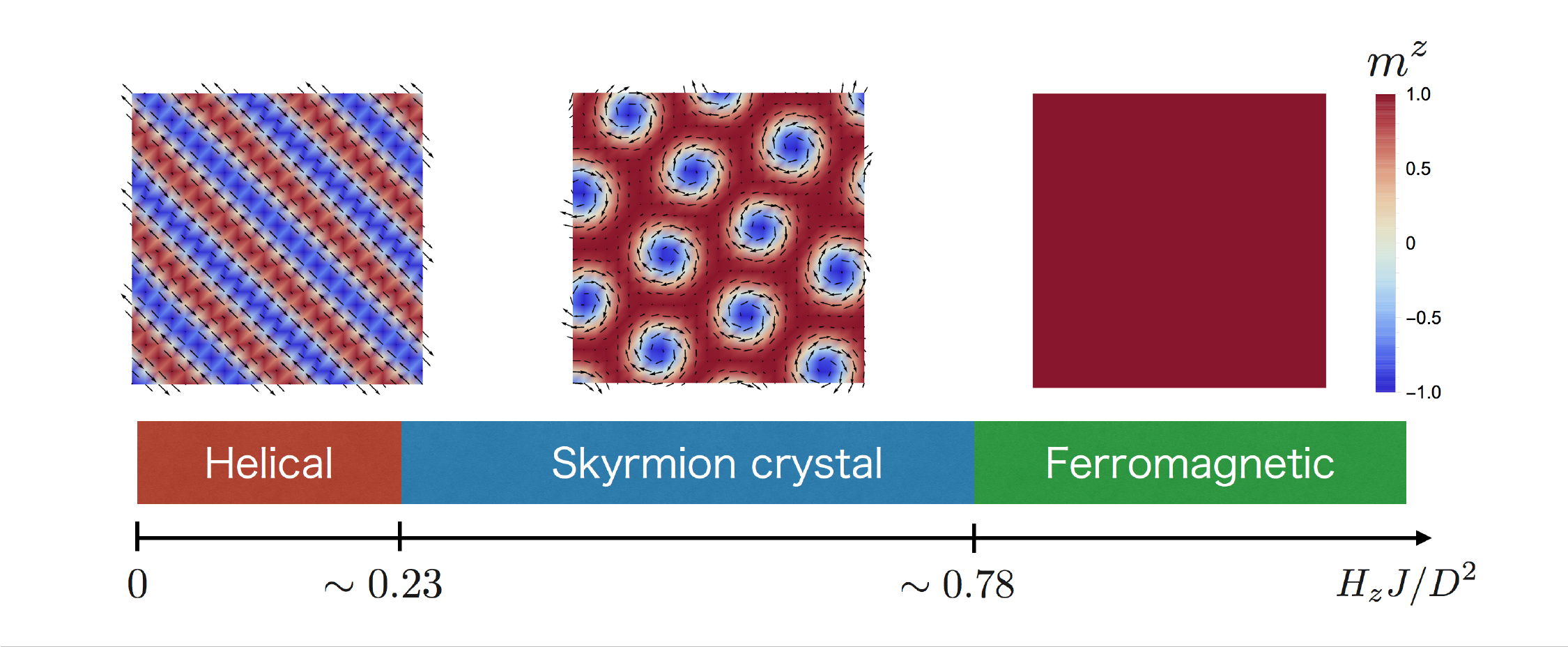}
       \caption{Ground state phase diagram of the model Eq.~\eqref{model}. Depending on the ratio $H_z J/D^2$, we have three distinct phases. We show typical spin configurations in these phases obtained by LLG calculations. The color indicates the $z$ component of spins and arrows do their $x$, $y$ components.}
          \label{phasediag_chiralFM} 
  \end{figure}

In the main text, we used a canonical model of two dimensional chiral ferromagnets on a square lattice:
\begin{align}
H = - J \sum_{\vec{r}} \vec{m}_{\vec{r}}\cdot\left(\vec{m}_{\vec{r} + a \vec{e}_{x}}+ \vec{m}_{\vec{r} + a \vec{e}_{y}} \right)-H_z\sum_{\vec{r}}m^{z}_{\vec{r}}+\sum_{\vec{r},i}\vec{D}_{i}\cdot\left(	\vec{m}_{\vec{r}}\times \vec{m}_{\vec{r}+a\vec{e}_{i}}		 \right).
\label{model}
\end{align}
Here we shortly review the static properties of this model, particularly focusing on topological defects, skyrmions in that. The first term ($J>0$) describes the ferromagnetic Heisenberg interaction and the second term does the Zeeman coupling between magnetic moments and the external magnetic field $\vec{H} = (0, 0, H_z)$.
The third term, called the Dzyaloshinskii-Moriya (DM) term favors the twisted spin texture. These three terms compete with each other and result in the ground state phase diagram shown in Fig.~\ref{phasediag_chiralFM}.
There appear three distinct phases depending on the ratio $H_z J/D^2$ (here  $\vec{D}_{i} = D \vec{e}_{i}$). Under the strong external magnetic field $H_z$, the system is ferromagnetic and all the spins align in the positive $z$-direction. On the other hand, if the magnetic field is very weak, ferromagnetic Heisenberg interaction and DM interaction turn the system into the helically ordered state. In between them, there exists the skyrmion crystal phase where skyrmions, localized magnetic defects with (clockwise) in-plane spin configuration, form a triangular lattice in the ground state. Skyrmions can also appear in the form of isolated magnetic defects in the ferromagnetic phase as metastable defects.

\begin{figure}[htbp]
   \centering
\includegraphics[width = 120mm]{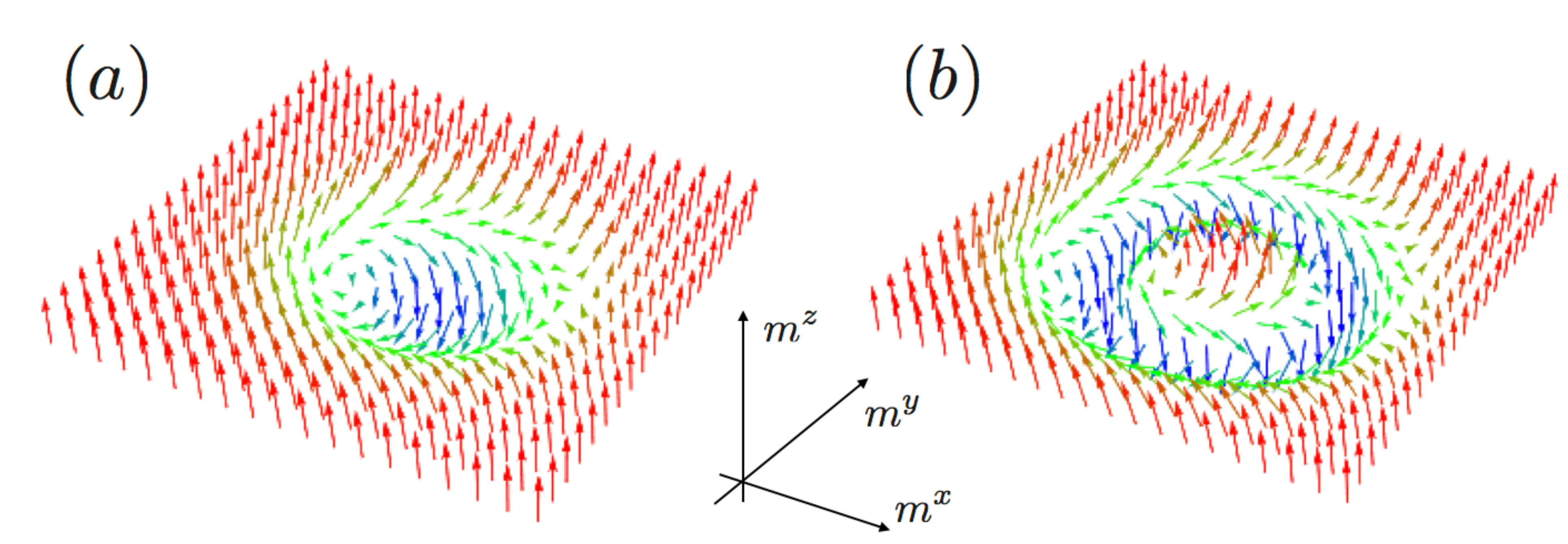}
       \caption{Spin structure of a ($a$) skyrmion and ($b$) skyrmionium in the ferromagnetic background with $m^z = +1$.}
          \label{skyrmionic_defects} 
  \end{figure}

The spin configuration of a skyrmion is presented in Fig.~\ref{skyrmionic_defects}($a$) (and the middle panel of Fig.~\ref{phasediag_chiralFM}). Skyrmions are characterized by their nontrivial topological number $N_{\rm sk}$ defined by 
\begin{align}
N_{\rm sk} = \frac{1}{4\pi}\int \vec{m}_{\vec{r}}\cdot \left(\frac{\partial \vec{m}_{\vec{r}}}{\partial x}\times \frac{\partial \vec{m}_{\vec{r}}}{\partial y} \right)d^{2}r.
\label{skyrmion_num}
\end{align}
In the continuum space, Eq.~\eqref{skyrmion_num} is quantized to an integer and any smooth deformations to the spin texture cannot change $N_{\rm sk}$. Since the spin texture of skyrmions give a non-vanishing $N_{\rm sk}$, they are stable against continuous deformations. Although this quantization is imperfect in lattice systems, the topological nature of skyrmions makes them energetically meta-stable magnetic defects.

At the center of each skyrmion, spins point to the opposite direction with respect to the background spin structure. In other words, spins in skyrmions are $\pi$-rotated from those in the outside. Then, we can easily generalize this into $n\pi$-rotation. If $n = 2$, spins in and outside that defects point to the same direction ($+z$-direction for example), but they are separated by a closed domain formed by spins with negative $m^z$. As a whole, the defect with $2\pi$ spin rotation, called skyrmionium, looks like a ring in terms of $m^z$. In Fig.~\ref{skyrmionic_defects}, we show the spin texture of a $(a)$ skyrmion ($\pi$ vortex) and $(b)$ skyrmionium ($2\pi$ vortex). 

\section{Use of beams with larger waist} \label{larger}
In the main text, we assumed the deep subwavelength focusing of optical vortices. In this section, we show that this assumption is not quite essential for our results, i.e. we can reproduce the qualitatively the same results using optical vortices with much larger beam waist for proper parameters.

\begin{figure}[htbp]
   \centering
\includegraphics[width = 100mm]{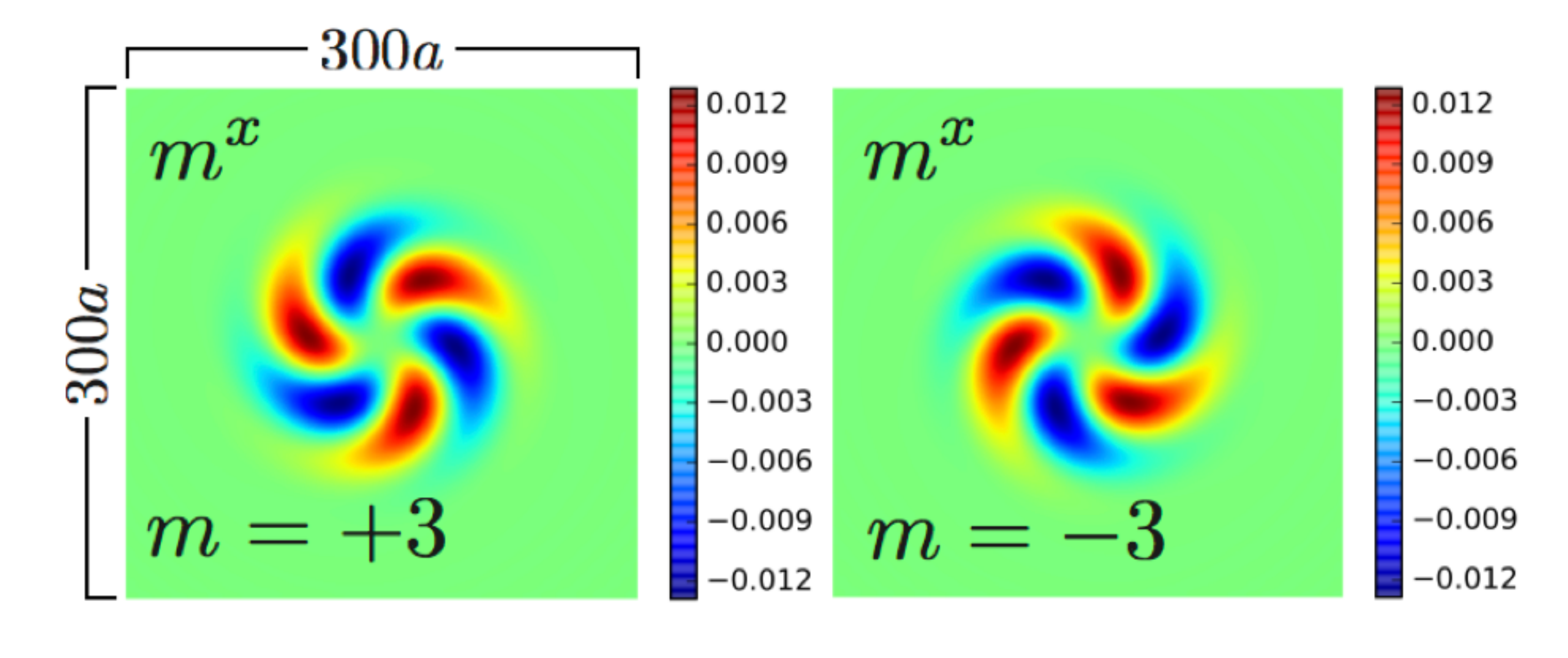}
       \caption{Spiral spin wave excited by linearly polarized beam $\vec{e}_p = \hat{x}$ at the resonance $\omega = H_z  = 0.3$. We apply beams with $W = 30 a$, $B_0 = 0.1$, and $m = \pm 3$ to the perfect ferromagnetic state ($m^z_{\vec{r}} = 1$ for all sites $\vec{r}$) of the model without DM interaction $D = 0$. The temporal profile of the pulse is the same as Fig.~3 in the main text. We show snapshots of the $x$-component of spins at $t = 100$. The system size is $300a$ times $300a$ where $a$ is the lattice constant. }
          \label{30a_wave} 
  \end{figure}

In Fig.~\ref{30a_wave}, we show the anisotropic spin wave excitation induced by optical vortices with $W = 30 a$, which is four times larger than that assumed in the main text (the temporal profile of the beam is the same as in the main text Fig.~3). We see that the OAM dependence of the shape of the spin wavefronts is the same as in the main text. This is because there is no length scale in our model other than the lattice constant ($a$) (here we assume $D = 0$). Consequently as long as the beam waist of the optical vortex is much larger than the lattice constant, the shape of wavefronts induced by the beam becomes at least qualitatively the same even for different beam waists.

As for the generation of topological defects, however, matching the length scale of the beam waist and that determined by the DM interaction, i.e. the skyrmion size, is important. Although it is quite challenging to achieve the deep focusing down to nanometer scale, as we mentioned in the main text, the proper beam waist of optical vortex will be more accessible values by using materials with large skyrmions. 

\begin{figure}[htbp]
   \centering
\includegraphics[width = 100mm]{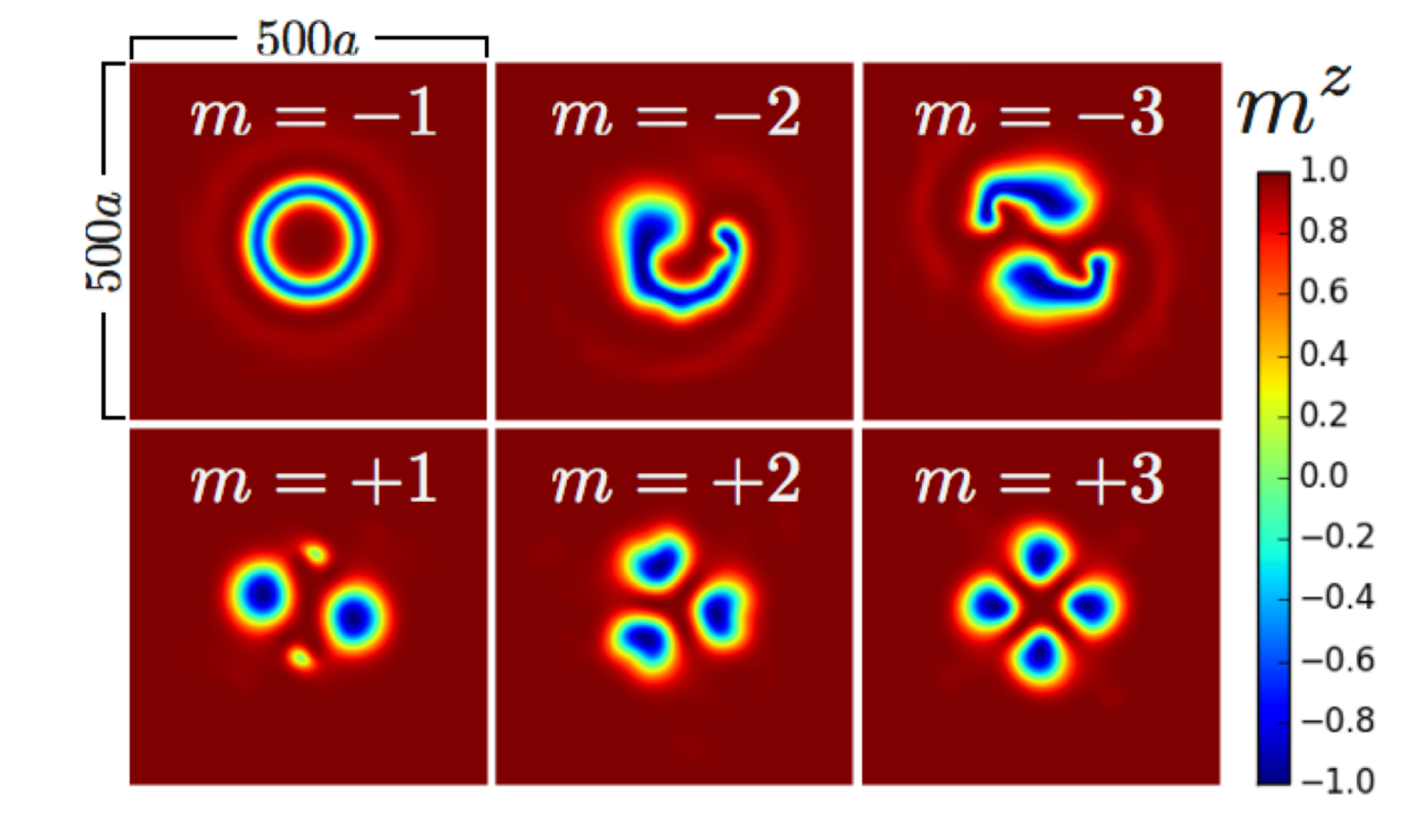}
       \caption{Transfer of orbital angular momentum from circularly polarized ($\vec{e}_p = \hat{x} + i \hat{y}$) optical vortices with $\omega = 0.075$, $B_0 = 0.15$, and $W = 50 a$ to the model of chiral magnets Eq.~\eqref{model} with $D = 0.07$, $H_z = 0.003$, and $\alpha = 0.1$. We show the spin structure at $t = 1500$, which is about 200 ps if $J = 5$ meV. The temporal profile of the pulse is the same as Fig.~4 in the main text. The system consists of 500 times 500 sites with the periodic boundary condition imposed.}
          \label{50a} 
  \end{figure}

As Fig.~\ref{50a} shows, by using smaller $D/J$ and $H_z/J$, or a model with larger skyrmions, indeed we can reproduce the results in Fig.~4 of the main text. Namely, the relationship between $m$ and the topological number after the irradiation for optical vortex in the main text holds for larger beam waist (here we assume $W = 50a$, five times larger than that in the main text). The important point is that we could obtain Fig.~\ref{50a} by using the optical vortex with the same temporal profile (a half-cycle pulse with the same frequency) as in Fig.~4 of the main text. That is, the proper frequency to cause this phenomenon, i.e. the OAM encoding, does not change even if we decrease $D$, or use materials with larger skyrmions. Therefore, from Fig.~\ref{50a} we can expect that the proper beam waist for THz optical vortex scales with the size of skrymions and the strong nano-focusing required to realize OAM encoding can be relaxed to a micrometer scale in reality.

\begin{figure}[htbp]
   \centering
\includegraphics[width = 180mm]{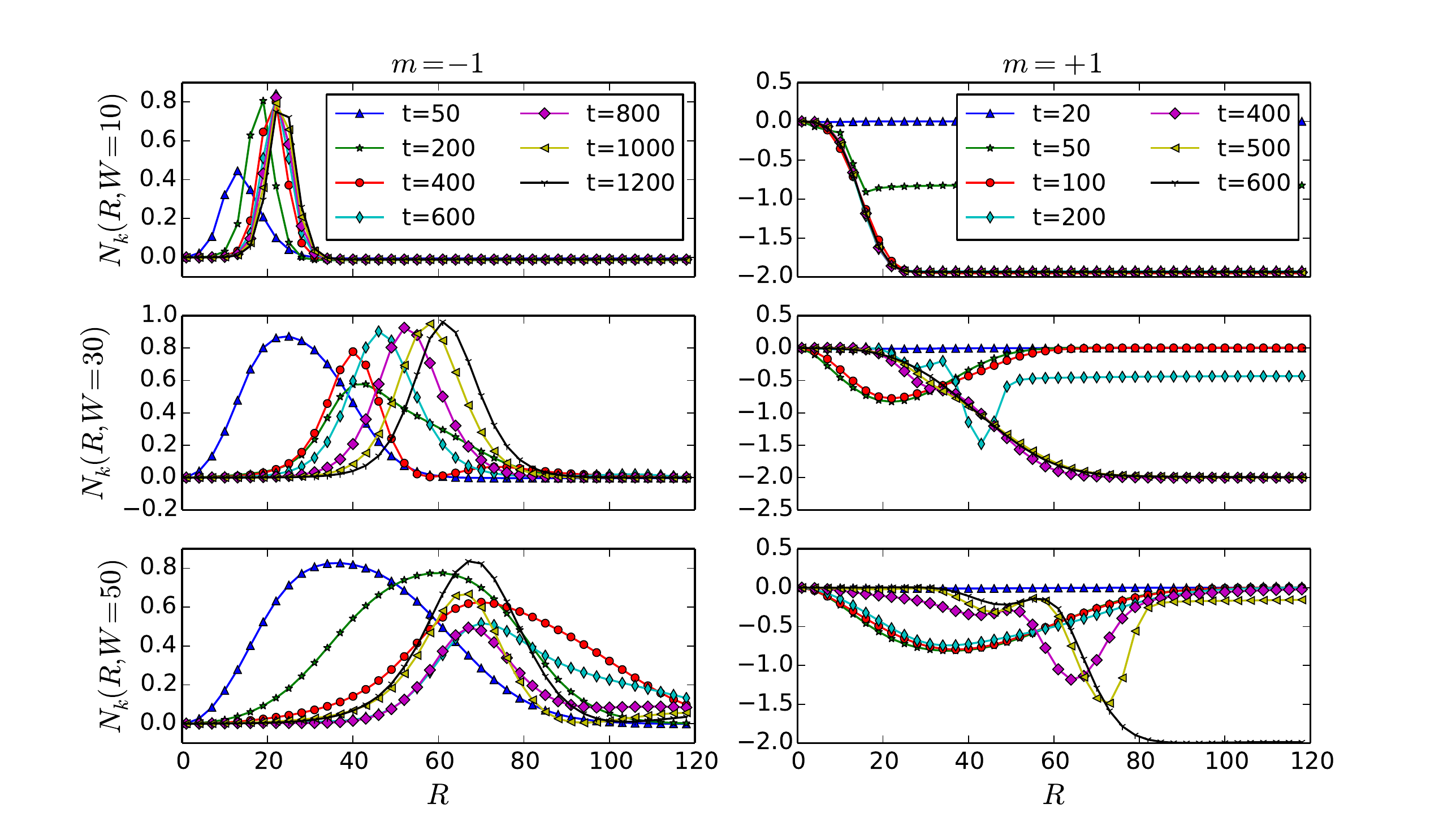}
       \caption{Time evolution of the cummulative skyrmion number for $m=\pm 1$ and $ W = 10a, 30a, 50a$. We fix $JH_z/D^2 =0.6$ and take $D = 0.08\frac{50a}{W}$. For small $D$ (large $W$), it takes longer time for the induced defects to relax to their equilibrium states. However, when $m = -1$, the characteristic bump structure of skyrmioniums is formed even right after the pulse passes, regardless of the beam waist $W$ (or the magnitude of $D$ equivalently). The radius $R$ is measured in the unit of $a$. }
          \label{TE_ws} 
  \end{figure}

Finally we discuss the timescale of the formation process.  In Fig.~\ref{50a} we show the snapshot for $W = 50 a$ at $t = 1500$, which are both five times larger and longer than those in the main text. For materials with larger skyrmions, it takes longer time for the laser-induced defects to be relaxed to their equilibrium states. However, we find that for $m = -1$, the topological structure itself develops at the very early stage of the equilibration process as shown in Fig.~\ref{TE_ws}. Here we plot the ``cummulative" skyrmion number
\begin{align}
N_k(R) = \frac{1}{4\pi}\int_{|\vec{r}| < R} \vec{m}_{\vec{r}}\cdot \left(\frac{\partial \vec{m}_{\vec{r}}}{\partial x}\times \frac{\partial \vec{m}_{\vec{r}}}{\partial y} \right)d^{2}r,
\label{cum_skyrmion_num}
\end{align}
calculated for spins within the circle with radius $R$ (measured from the center of the beam spot) for $m=\pm1$ and $W = 10a,$ $30a$, $50a$.  We fix $J H_z/D^2 = 0.6$ and take $D = 0.08\frac{50a}{W}$. Since the skyrmionium is a bound state of a skyrmion and anti-skyrmion, its cummulative skyrmion number shows the bump structure as a function of $R$. In Fig.~\ref{TE_ws}, we see that the formation of the bump is qualitatively completed even right after the optical vortex pulse passes (we are assuming a pulse with $\sigma = 10$, $t_0 = 30$) even when the beam waist (thus the size of skyrmion) is large. Therefore, the timescale of generating skyrmioniums with $m = -1$ optical vortices is essentially unchanged even for materials with large skyrmions. For $m\neq-1$, on the other hand, as we mentioned in the main text, the formation of the laser-induced defects depends strongly on the spin dynamics governed by the DM interaction. As a result, we see that the timescale becomes longer for larger beam waist. In the figure, as an example we show the results for $m = +1$ beams under which a pair of skyrmions (each has the skyrmion number $-1$) are created, and the total topological number after the laser irradiation is $-2$.
 
With increasing the size of skyrmions (and skyrmioniums) in the steady state, their creation/annihilation take longer time under the conventional schemes since their timescale is determined by the magnitude of $D$, which is smaller for large skyrmion materials. Our results show that at least for the creation (and possibly the annihilation) of skyrmioniums, optical vortices could resolve the timescale issue. Hence, as far as the topological structure is concerned, by using the subwavelength optical vortices,  we can generate the large skyrmioniums in the THz timescale.

\section{Magneto-electric coupling} \label{ME}
In insulating chiral magnets like Cu${}_2$OSeO${}_{3}$, magneto-electric (ME) coupling allows us to control the magnetic texture with electric fields, instead of magnetic fields. Because the electric field of a laser beam is usually much larger than the magnetic field, in such multiferroic materials with the large ME coupling constant, the magnetization dynamics can be dominated by the coupling of magnets with the electric fields. The ME coupling we discuss here is described by the following Hamiltonian:
\begin{align} 
H_{\mathrm{ME}} = -\sum_{\vec{r}} \vec{E}(\vec{r},t)\cdot \vec{p}(\vec{r},t).
\end{align}
In multiferroic materials, the local electric polarization $\vec{p}(\vec{r})$ is a function of local spins.
The electric field at the focal plane $z = 0$ is $\vec{E}(\vec{r}, t) = \vec{e}_{p}E(\vec{r}, t)$ with polarization vector $\vec{e}_{p}$ and
\begin{align}
 E(\vec{r}, t) = \frac{|E_0|E_{m, p}(\vec{r})}{\max_{\vec{r}}|E_{m, p}(\vec{r})|}\exp \left[-\left(\frac{t - t_0}{\sigma}\right)^2 - i \omega t \right].
 \label{elfield}
 \end{align}
Here $E_{m, p}(\rho,\phi,0)$ is the LG mode of electric field with the radial index $p$ and OAM $m$.

In the following, we discuss two major mechanisms of ME coupling in multiferroic magnets: Dzyaloshinskii-Moriya (DM) interaction induced coupling~\cite{PhysRevLett.95.057205} and $p$-$d$ hybridization induced one~\cite{Arima:2007aa,PhysRevB.86.060403}. The former mechanism gives the electric polarization proportional to the outer-product of two nearby spins: $\vec{p}(\vec{r}_{i,i+1}) \propto \vec{e}_{i, i+1}\times (\vec{m}_{\vec{r}_i}\times\vec{m}_{\vec{r}_{i+1}})$, where $\vec{e}_{i,i+1}$ is the unit vector pointing from site $\vec{r}_i$ to site $\vec{r}_{i+1}$. Finite polarization from this mechanism requires a canted magnetic structure and does not exist in the collinear magnetic states. Therefore, even if the coupling constant between spins and electric field is large, as long as we are considering of sufficiently smooth magnetic textures like in spin waves or skyrmions, ME coupling of this type becomes effectively weaken and Zeeman coupling discussed in the main text would dominate the response. 

On the other hand, ME coupling from $p$-$d$ hybridization could be important in our setup. The ME coupling of this kind is experimentally observed in Cu${}_2$OSeO${}_{3}$, and described by the polarization vector $\vec{p}(\vec{r}) =\lambda (m^y_{\vec{r}}m^z_{\vec{r}},m^z_{\vec{r}}m^x_{\vec{r}},m^x_{\vec{r}}m^y_{\vec{r}})$~\cite{AELM:AELM201500180}, if we take the $c$-axis to be the $z$-axis. For Cu${}_2$OSeO${}_{3}$, experimentally we have $\lambda = 5.64 \times 10^{-33}$ Cm~\cite{PhysRevB.86.060403, AELM:AELM201500180}. Contrary to the DM induced one, this ME coupling can exist in collinear magnetic phase and can dominate over the effect of Zeeman coupling. Substituting this polarization vector into the Hamiltonian $H_{\mathrm{ME}}$ and differentiate it by $-\vec{m}_{\vec{r}}$, we obtain the following effective magnetic field from the ME coupling, which enters in the LLG equation as
\begin{align}
\vec{H}^{\mathrm{ME}}_{\mathrm{eff}} = \lambda \left(E^ym^z_{\vec{r}} ,E^xm^z_{\vec{r}}, E^xm^y_{\vec{r}} +E^ym^x_{\vec{r}}\right).
\label{ME_ham}
\end{align}
Then we notice that the in-plane components of the effective field are proportional to $m^z_{\vec{r}}$. Therefore, the torque which is to rotate spins from the initial state $\{m^z_{\vec{r}}\} = 1$ gets weaker and weaker as $m^z_{\vec{r}}$ approaches to zero. As a result, the precession induced by ME coupling tends to confine spins inside $x$-$y$ plane rather than to induce magnetization reversal. 

\begin{figure}[htbp]
   \centering
\includegraphics[width = 160mm]{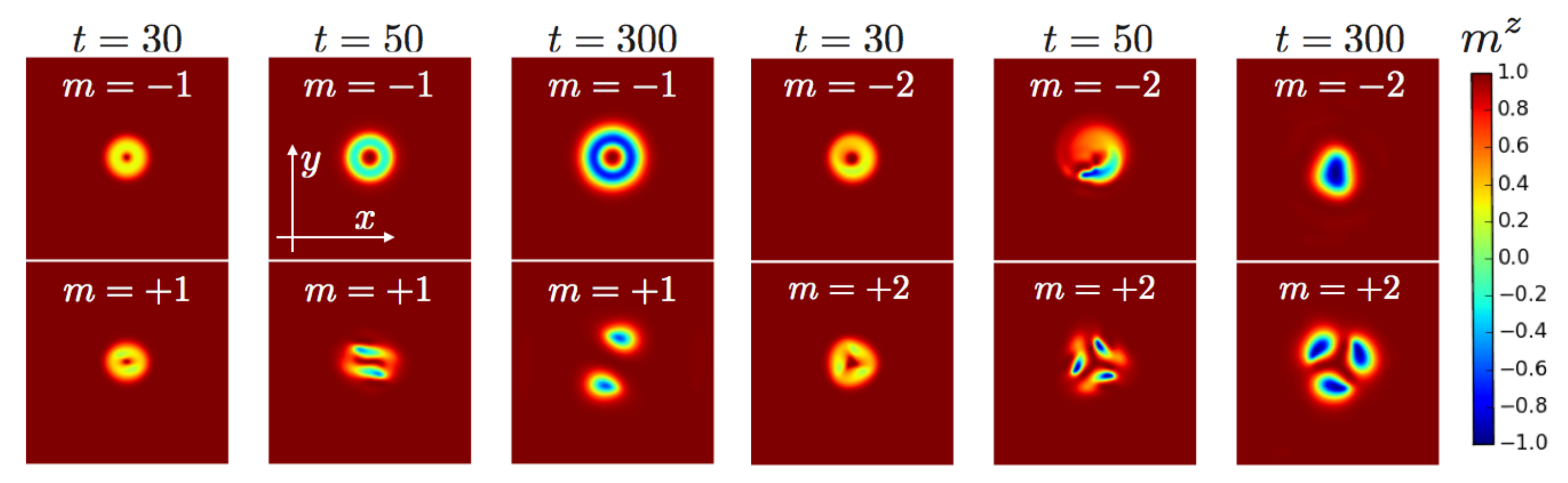}
       \caption{Orbital angular momentum dependent creation of topological defects by magneto-electric coupling between spins and right-handed optical vortices with $\vec{e}_p = \hat{x} - i \hat{y}$. We present time evolution of spins for $D = 0.15$, $H_{z} = 0.015$, $W= 10 a$, $p = 0$, $\omega = 0.075$, $\sigma = 10$, $t_0 = 30$, $E_0 \lambda = 0.25$, and $\alpha = 0.1$. The system is periodic in both $x$ and $y$ directions with 150 times 150 sites.}
          \label{multiferro} 
  \end{figure}

Even though the precession itself is not completely suitable for the creation of topological defects, it still twists the spin texture and could generate topological defects with the help of DM and Heisenberg interaction. With numerical calculations based on LLG equation for the model:
\begin{align}
H &= - J \sum_{\vec{r}} \vec{m}_{\vec{r}}\cdot\left(\vec{m}_{\vec{r} + a \vec{e}_{x}}+ \vec{m}_{\vec{r} + a \vec{e}_{y}} \right)-H_z\sum_{\vec{r}}m^{z}_{\vec{r}} +\sum_{\vec{r},i}\vec{D}_{i}\cdot\left(	\vec{m}_{\vec{r}}\times \vec{m}_{\vec{r}+a\vec{e}_{i}}		 \right) + H_{\mathrm{ME}},
\label{model_S}
\end{align}
we find that for right-handed optical vortices with $\vec{e}_{p} = \hat{x} - i \hat{y}$, the ME coupling from $p$-$d$ hybridization Eq.~\eqref{ME_ham} can indeed generate topological defects just as the Zeeman coupling with left-handed optical vortices. In Fig.~\ref{multiferro}, we show time evolution of spins for $D = 0.15$, $H_{z} = 0.015$, and $\alpha = 0.1$ under the electric field Eq.~\eqref{elfield} with $W = 10 a$, $p = 0$, $\omega = 0.075$, $\sigma = 10$, $ t_0 = 30$, and $E_0 \lambda = 0.25$. For $J \sim 1$ meV and $\lambda \sim 10^{-32}$ Cm, the electric field strength corresponds to $40$ MV$/$ cm. We see that the final states after the irradiation are qualitatively the same as those in the case of Zeeman coupling discussed in the main text. Namely, we can generate a skyrmionium and skyrmions depending on the OAM of optical vortices. For left-handed electric field, we do not observe any topological defects after the irradiation for all these OAMs.

\begin{figure}[htbp]
   \centering
\includegraphics[width = 90mm]{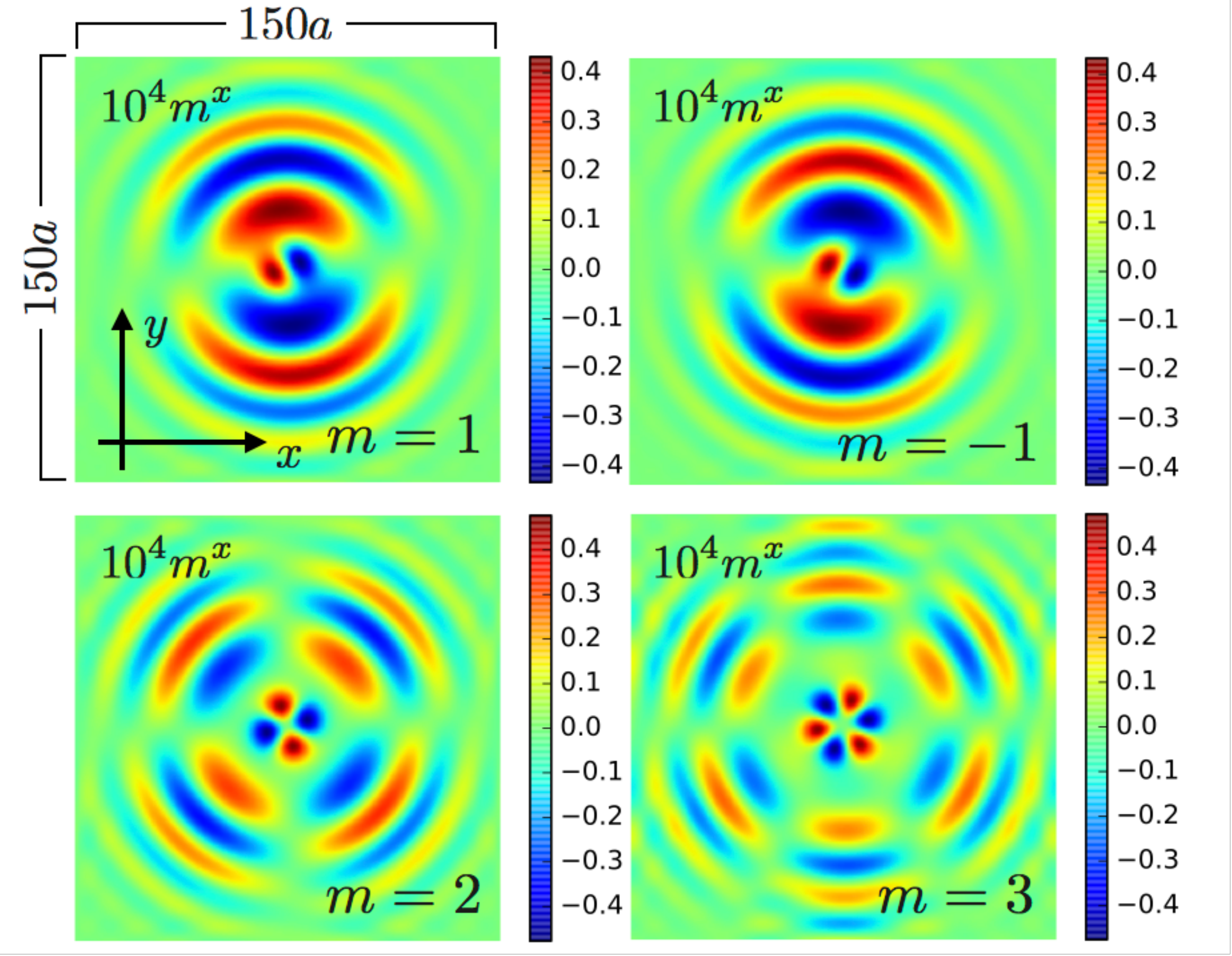}
       \caption{Multipolar spin wave induced by magneto-electric coupling with linearly polarized ($\vec{e}_{p} = \hat{x}$) optical vortices Eq.~\eqref{elfield} for $D = 0$, $H_{z} = 0.015$, $W = 7.5 a$, $p = 0$, $\omega = 2$, $\sigma = 20$, $t_0 = 40$, $E_0 \lambda = 0.05$, and $\alpha = 0.1$. We show snapshots of the $x$-component of spins at $t=90$ for several values of $m$.}
          \label{sw_EM} 
  \end{figure}

We also observe multipolar spin wave excitations for linearly polarized electric fields with $\vec{e}_p = \hat{x}$ as shown in Fig.~\ref{sw_EM}, just as those from Zeeman coupling discussed in the main text. Here we assume $\lambda E_0 = 0.05$ and $\omega = 2$. For $J = 1$ meV, $E_0 = 8$ MV$/$cm and $\omega = 0.5$ THz. In this setup, even under the laser irradiation, $m^z \simeq 1$, $m^x\simeq 0$, and $m^y\simeq 0$ are preserved since the electric field field is rather weak. Therefore the effective field Eq.~\eqref{ME_ham} is almost equivalent to Zeeman coupling with a magnetic field $\vec{B} = (\lambda E^y, \lambda E^x,0)$, so it is natural to have qualitatively the same outcome. The situation is the same at the resonance $\omega = H_z$ and we have spiral-shaped spin waves with larger amplitudes.

\section{Gaussian beam}\label{Gaussian}
\begin{figure}[htbp]
   \centering
\includegraphics[width = 90mm]{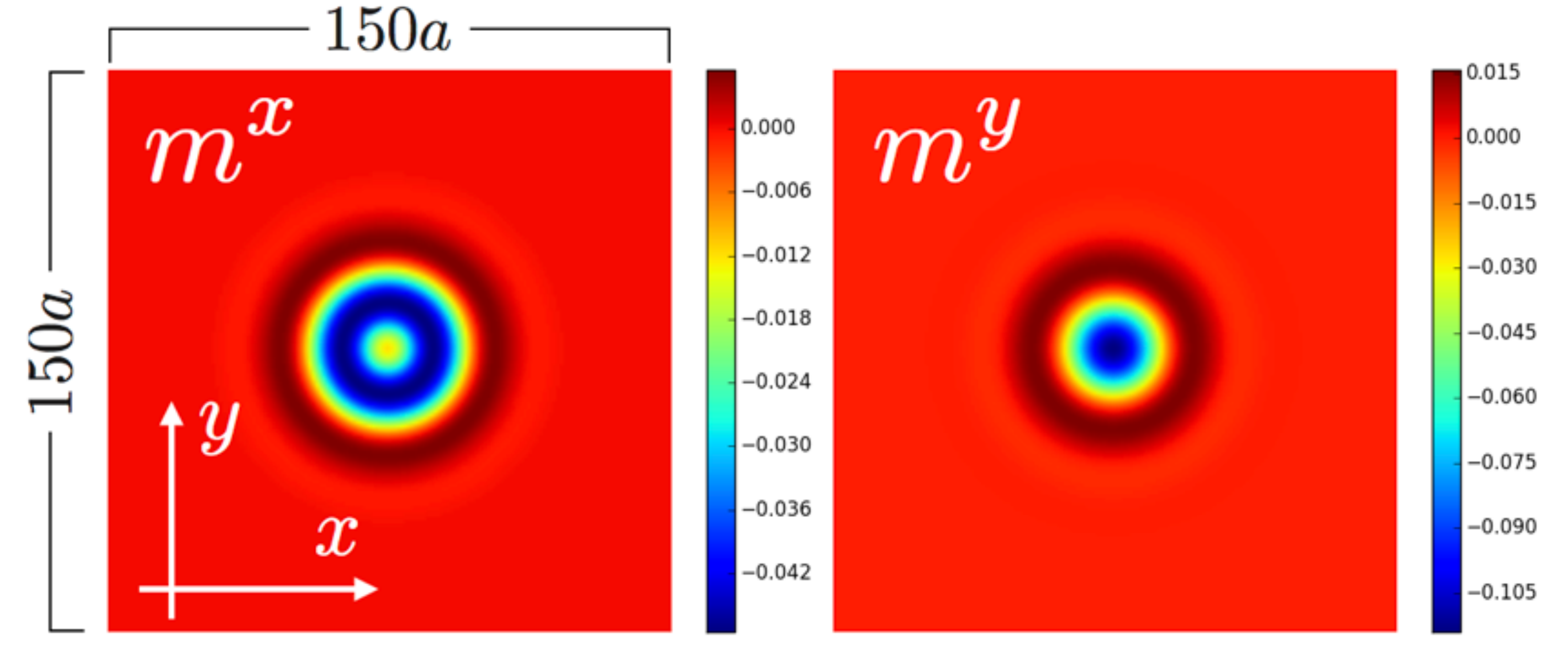}
       \caption{Isotropic spin wave excitation by linearly polarized Gaussian beam with $m = 0$ for $\vec{e}_{p} = \hat{x}$, $D = 0, \omega = H_{z} = 0.3, W = 7.5 a, p = 0, B_0 = 0.05$, $\sigma = 20$, $t_0 = 40$, and $\alpha = 0.1$. The spin structure at $t = 80$ is shown. }
          \label{m=0_sw} 
  \end{figure}

\begin{figure}[htbp]
   \centering
\includegraphics[width = 130mm]{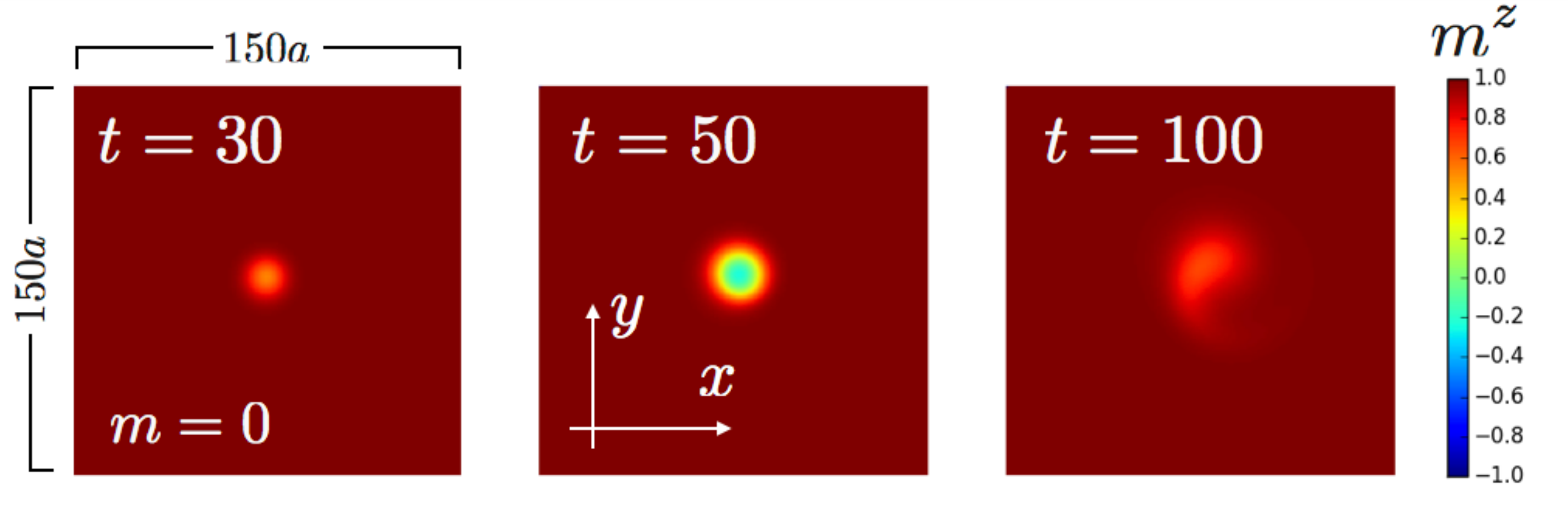}
       \caption{Irradiation of Gaussian beam magnetic field without orbital angular momentum for $H_{z} = 0.015$, $W = 10 a$, $p = 0$, $\omega = 0.075$, $\sigma = 10$, $t_0 = 30$, $B_0 = 0.15$, $\alpha = 0.1$, and $\vec{e}_{p} = \hat{x} + i \hat{y}$, same as the corresponding calculation for optical vortices (finite $m$ beams) in the main text. Because Gaussian beam with $m=0$ does not add twist to the spin structure, it is not effective for the creation of topological defects. }
          \label{m=0_mionium} 
  \end{figure}

In the main text, we saw that linearly polarized optical vortices with non-vanishing OAM can induce spin dipolar-, quadrupolar-, and octapolar- radiations, and left-handed, circularly polarized optical vortices offer an efficient way of creating topological defects in chiral magnets. Here for comparison we show the result for $m = 0$, namely ordinary Gaussian beams without carrying OAM. As shown in Fig.~\ref{m=0_sw}, even for the linearly polarized $\vec{e}_p$ Gaussian beam at the resonance, the spin wave is excited, but its spatial distribution is radially isotropic, contrary to the multipolar radiations from finite OAM lasers discussed in the main text. Since the $m=0$ Gaussian beam does not twist the spin texture non-uniformly unlike optical vortex, as shown in Fig.~\ref{m=0_mionium}, we could not obtain any topological defects after the irradiation for left-handed beam even with the same beam strength used in the main text.
  
\begin{figure}[htbp]
   \centering
\includegraphics[width = 125mm]{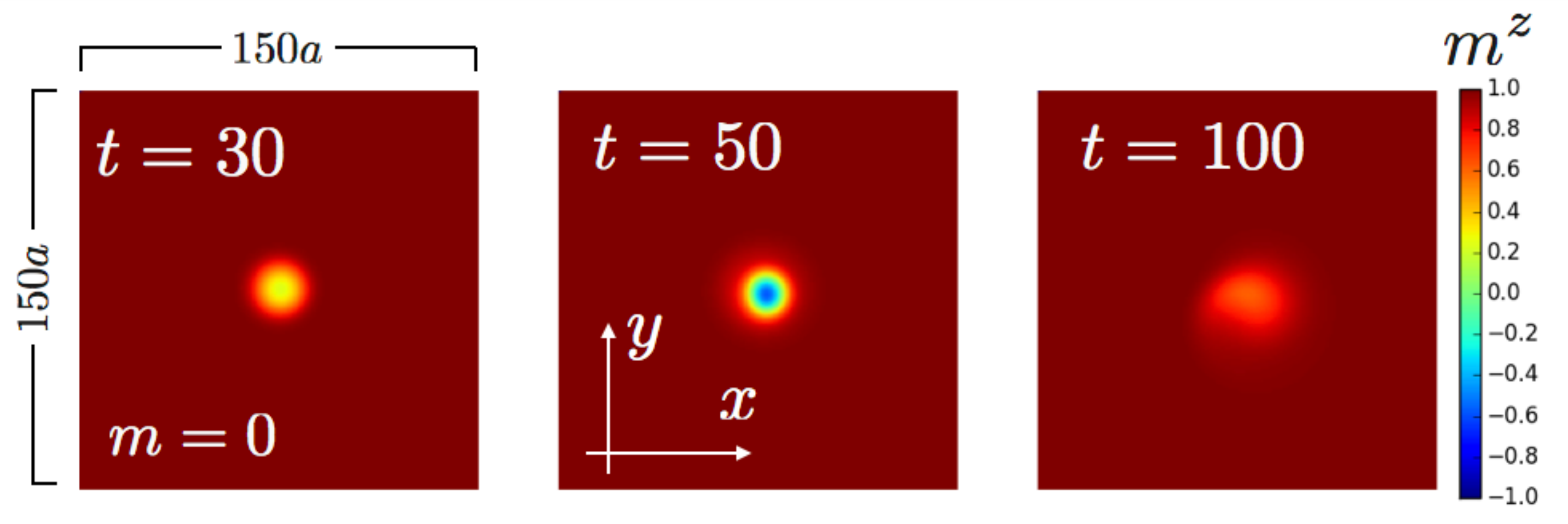}
       \caption{Irradiation of Gaussian beam electric field without orbital angular momentum for $H_{z} = 0.015$, $W = 10 a$, $p = 0$, $\omega = 0.075$, $\sigma = 10$, $t_0 = 30$, $E_0\lambda = 0.25$, $\alpha = 0.1$, and $\vec{e}_{p} = \hat{x} - i \hat{y}$, same as the corresponding calculation in Fig.~\ref{multiferro}.}
          \label{multiferro_m0} 
  \end{figure}

 The situation is the same for ME coupling as shown in Fig.~\ref{multiferro_m0}. That is, Gaussian beams without OAM are not suitable for creating topological defects unlike optical vortices in Fig.~\ref{multiferro}.

 \section{Optical polarization dependence}\label{polarization_dep}
 Here we discuss the optical polarization dependence of the phenomena discussed in the main text, namely spin waves and creation of topological defects.
  \subsection{Anisotropic spin waves}
\begin{figure}[htbp]
   \centering
\includegraphics[width = 150mm]{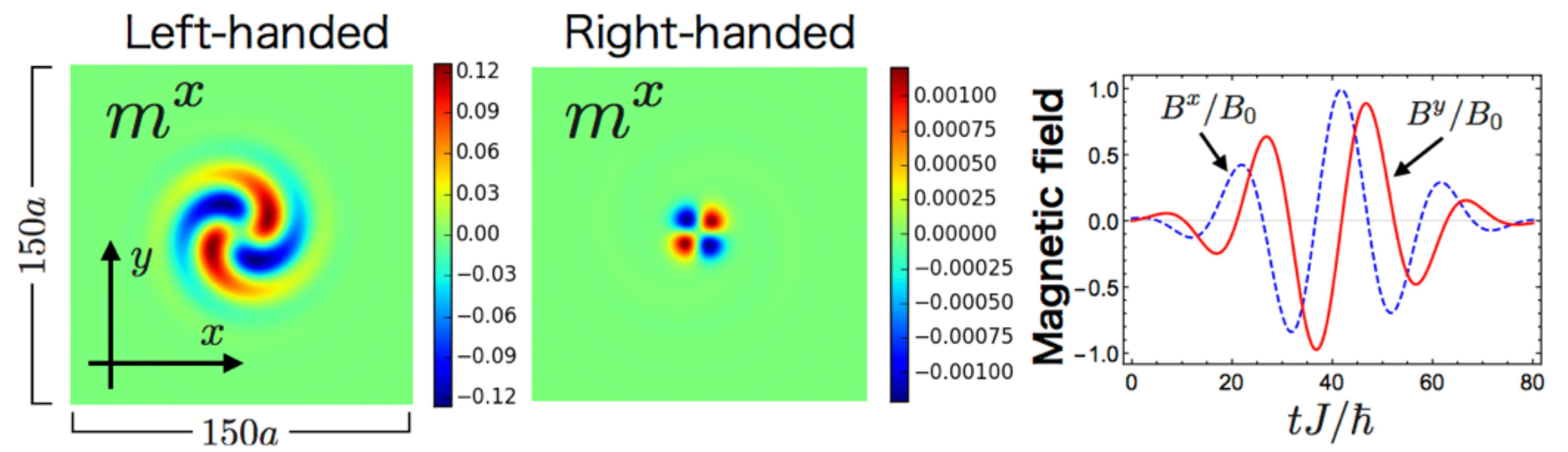}
       \caption{Spiral spin waves induced by circularly polarized magnetic fields with $m = 2$ for $D = 0$, $\omega = H_z = 0.3$, $B_0 = 0.05$, $\alpha = 0.1$, $W = 7.5 a$, $t_0 = 40$, and $\sigma = 20$. Snapshots at $t = 80$ are shown for both left and right handed beams with $\vec{e}_p = \hat{x} \pm i\hat{y}$. We also show the time-dependence of magnetic fields for the left handed case at $\phi = 0$ and $\rho = w$.  }
          \label{circular_res}  
  \end{figure}

  In the spin wave part of the main text, we focused on the linearly polarized beams $\vec{e}_p =\hat{x}$ and discussed multipolar spin waves from high frequency beams and spiral spin waves from beams with the frequency at the magnetic resonance. Here we discuss spin wave excitations at the magnetic resonance by circularly polarized beams. We consider both left and right handed beams $\vec{e}_p = \hat{x} \pm i\hat{y}$. For OAM $m = 2$, in Fig.~\ref{circular_res}, we show the $x$-component of spins after the laser irradiation. For the left handed cases, we clearly see the spiral-shaped wavefronts with the wave amplitude larger than the linearly polarized case in the main text. For the right handed cases, however, the amplitude is much smaller. The spatial structure of the wavefronts is spiral but unclear from the figure due to the small amplitude of the spin wave. The spiral wave fronts can be observed for the left-handed off-resonant beams with $\omega \agt J, H_z$, but the wave amplitude gets much smaller just as we saw for linearly polarized case in the main text.
  
 \subsection{Creation of topological defects}
\begin{figure}[htbp]
   \centering
\includegraphics[width = 100mm]{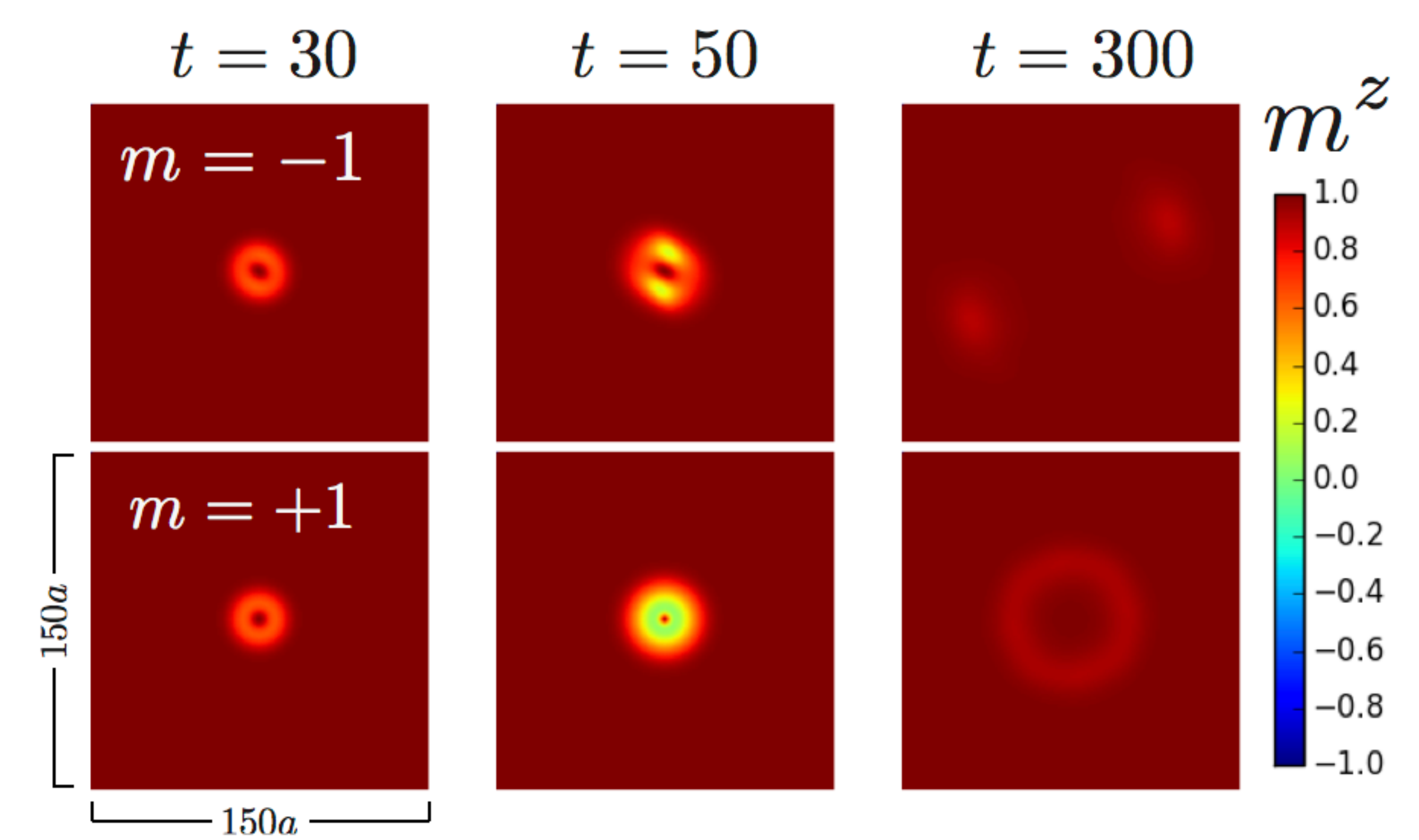}
       \caption{Irradiation of right-handed magnetic field with $\vec{e}_p=\hat{x} - i \hat{y}$ and orbital angular momentum $m = \pm1$. Time evolutions of the $z$ component of spins for $H_{z} = 0.015$, $W = 10 a$, $p = 0$, $\omega = 0.075$, $\sigma = 10$, $t_0 = 30$, $B_0 = 0.15$, $D = 0.15$, and $\alpha = 0.1$ are presented.}
          \label{negpol_defect} 
  \end{figure}

In the main text, we considered left-handed waves with $\vec{e}_{p} = \hat{x} + i \hat{y}$ to create magnetic defects. If we take $\omega = 0$, left-handed optical vortices with OAM $m >0$ and right-handed ones with OAM $-m$ are equivalent. However, once we have finite $\omega$, there appears clear difference. It turns out that right-handed optical vortex with $\vec{e}_{p} = \hat{x} - i \hat{y}$ is not suitable for the creation of topological defects as opposed to the left-handed one. We calculate the time evolution of spins for the right handed beams using the same parameters in the Fig.~4 of the main text: $H_{z} = 0.015$, $W = 10 a$, $m=\pm1$, $p = 0$, $\omega = 0.075$, $\sigma = 10$, $t_0 = 30$, $B_0 = 0.15$, $D = 0.15$, and $\alpha = 0.1$. As Fig.~\ref{negpol_defect} shows, with these parameters, we cannot get any topological defects, contrary to the left-handed. 

\begin{figure}[htbp]
   \centering
\includegraphics[width = 140mm]{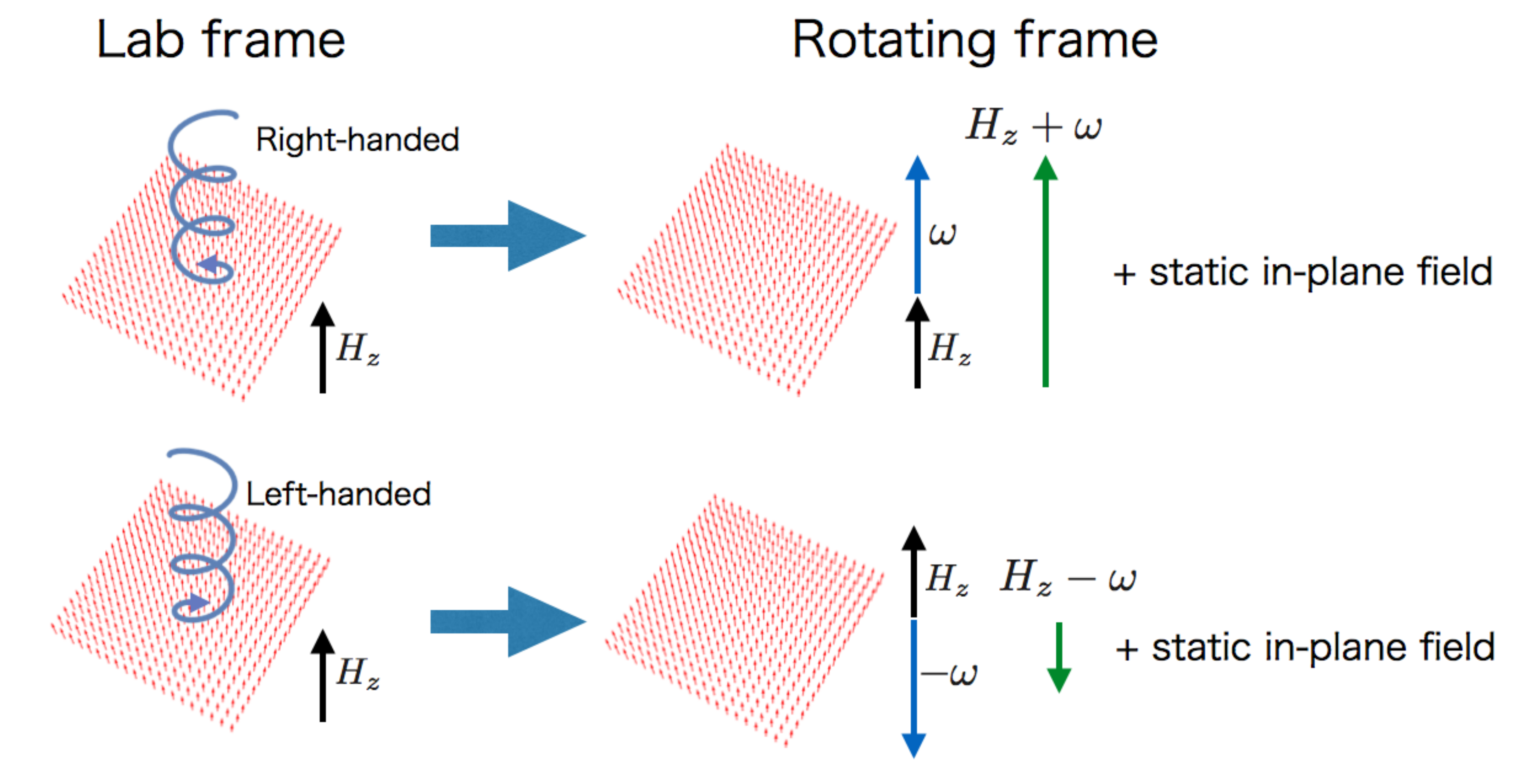}
       \caption{The rotating-frame interpretation of the polarization dependence. By moving into the rotating frame, we can transform the circularly polarized beam into a sum of the static in-plane magnetic field and the out-of-plane effective field.}
          \label{rotatingframe} 
  \end{figure}

We can understand the difference between left and right handed beams by moving into the rotating frame with frequency $\omega$ as shown in Fig.~\ref{rotatingframe}. In this frame, the magnetic field of optical vortex is transformed into a superposition of the static in-plane (anisotropic) magnetic field and the rotation-induced static magnetic field in the $z$-direction. The sign of the rotation-induced field depends on whether the original optical vortex is left or right handed. In our setup, rotation-induced field is in the $(-z)$-direction for left-handed optical vortex, which is the opposite to the uniform external magnetic field. Therefore, left-handed beams are more suitable for exciting spin waves and creating topological defects.

\begin{figure}[htbp]
   \centering
\includegraphics[width = 100mm]{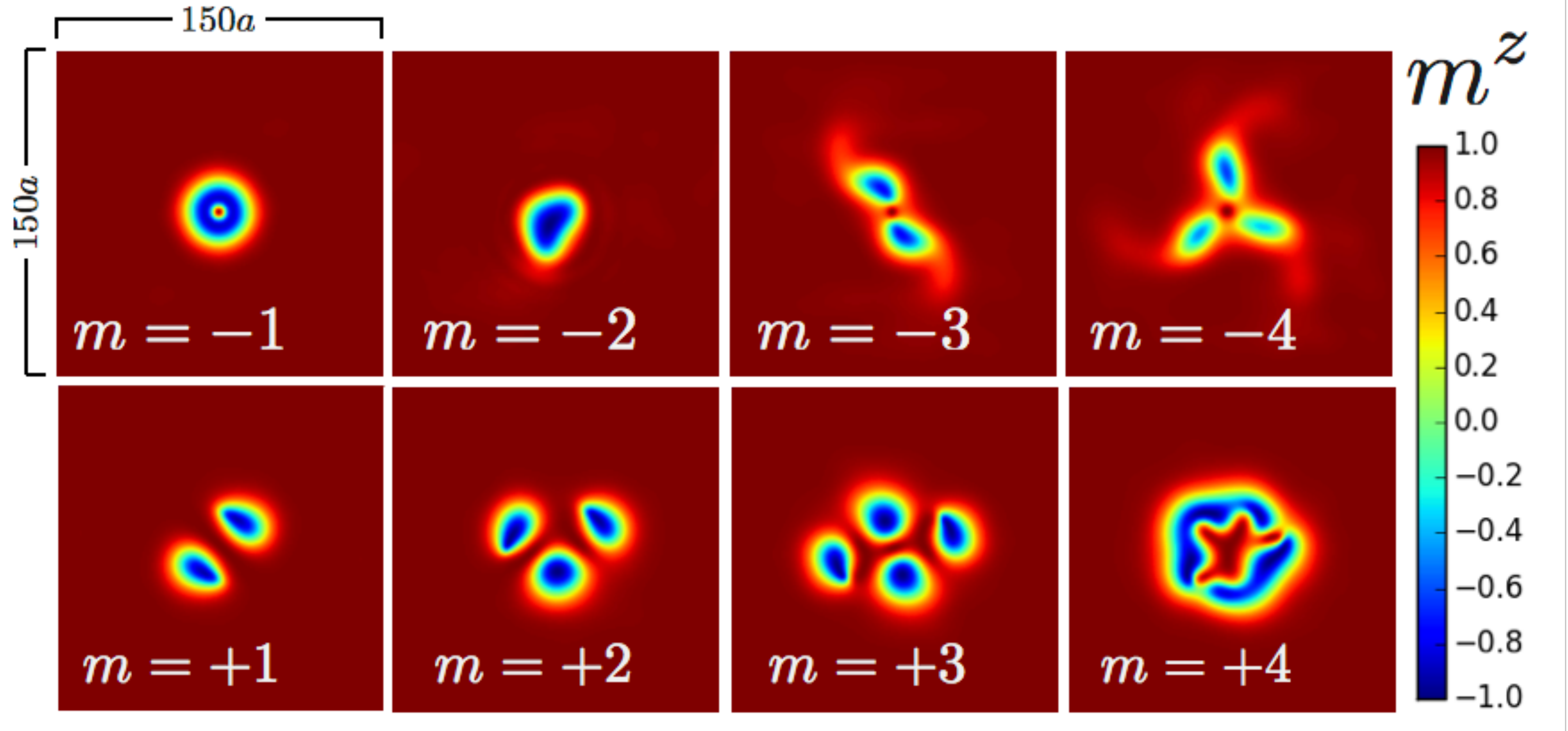}
       \caption{Defect creation with linearly-polarized magnetic field $\vec{e}_p=\hat{x}$ for several values of orbital angular momentum. We present snapshots of the spin textures at $t = 300$ for $H_{z} = 0.015$, $W = 10 a$, $p = 0$, $\omega = 0.075$, $\sigma = 20$, $t_0 = 40$, $B_0 = 0.15$, $D = 0.15$, and $\alpha = 0.1$.  }
          \label{linear_defect} 
  \end{figure}

We also comment on the use of linearly polarized magnetic fields for the creation of defects. Since linearly polarized field with, for example, $\vec{e}_p = \hat{x}$ is a superposition of left and right-handed beams, as a tool for creating defects, it would be more suitable than right-handed beams.  We calculated the time evolution for $H_{z} = 0.015$, $W = 10 a$, $p = 0$, $\omega = 0.075$, $\sigma = 20$, $t_0 = 40$, $B_0 = 0.15$, $D = 0.15$, and $\alpha = 0.1$ and summarize the result in Fig.~\ref{linear_defect}. Here the beam is still a half-cycle like but with the longer duration $\sigma = 20$. With $\sigma = 10$, that used in the circularly polarized cases in the main text, we could not obtain topological defects. We see from Fig.~\ref{linear_defect} that the linearly polarized laser indeed produces topological defects, though the systematic OAM dependence breaks down at $m=4$.

\end{document}